\documentclass[12pt,preprint]{aastex}

\newcommand{\subi}{_{i}}
\newcommand{\subim}{_{i-1}}
\newcommand{\subip}{_{i+1}}
\newcommand{\subj}{_{j}}
\newcommand{\subjm}{_{j-1}}
\newcommand{\subjp}{_{j+1}}

\newcommand{\DbDt}{{D \over Dt}}
\newcommand{\dbdt}{{d \over dt}}
\newcommand{\divf}{{\bf \nabla \cdot F}}
\newcommand{\chif}{\chi_{\bf F}}
\newcommand{\kp}{\kappa_{P}}
\newcommand{\ke}{\kappa_{E}}
\newcommand{\divv}{{\bf \nabla \cdot v}}
\newcommand{\gradvp}{{\bf \nabla v:P}}
\newcommand{\supn}{^{n}}
\newcommand{\supnp}{^{n+1}}

\newcommand{\F}{{\bf F}}
\newcommand{\p}{{\bf P}}
\newcommand{\f}{{\bf f}}
\newcommand{\divp}{{\bf \nabla \cdot \p}}
\newcommand{\deldot}{{\bf \nabla \cdot}}
\newcommand{\divfe}{{\bf \nabla \cdot} (\f E)}
\newcommand{\zmp}{ZEUS-MP}
\newcommand{\ztwd}{ZEUS-2D}
\newcommand{\ztwdft}{ZEUS2D-FT}
\newcommand{\ttwo}{$T_{2}$}
\newcommand{\tm}{$T_{-}$}
\newcommand{\afl}{_{\rm AFL}}
\newcommand{\itd}{_{\rm ITD}}
\newcommand{\subimj}{_{i-1,j}}
\newcommand{\subij}{_{i,j}}
\newcommand{\subipj}{_{i+1,j}}
\newcommand{\subimjm}{_{i-1,j-1}}
\newcommand{\subijm}{_{i,j-1}}
\newcommand{\subipjm}{_{i+1,j-1}}
\newcommand{\subimjp}{_{i-1,j+1}}
\newcommand{\subijp}{_{i,j+1}}
\newcommand{\subipjp}{_{i+1,j+1}}
\newcommand{\eav}{\langle E\rangle}
\newcommand{\bone}{{\rm b^{(1)}}}
\newcommand{\btwo}{{\rm b^{(2)}}}
\begin{document}

\title{Beyond Flux-Limited Diffusion: Parallel Algorithms
for Multidimensional Radiation Hydrodynamics}

\author{John C. Hayes and Michael L. Norman}
\affil{Physics Department, University of California,
    San Diego, CA 92093}

%

\begin{abstract}

This paper presents a new code for performing multidimensional
radiation hydrodynamic (RHD) simulations on parallel computers
involving anisotropic radiation fields and nonequilibrium effects.  The
radiation evolution modules described here encapsulate the physics
provided by the serial algorithm of~\citet{sto92c}, but add new
functionality both with regard to physics and numerics.
In detailing our method, we have documented both the analytic
and discrete forms of the radiation moment solution and the variable
tensor Eddington factor (VTEF) closure term.  We have described three different
methods for computing a short-characteristic formal solution to the 
transfer equation, from
which our VTEF closure term is derived.  Two of these techniques
include time dependence, a primary physics enhancement of the method
not present in the Stone algorithm.
An additional physics modification is the adoption of a matter-radiation
coupling scheme which is particularly robust for nonequilibrium
problems and which also reduces the operations cost of our radiation
moment solution. Two key numerical
components of our implementation are highlighted: the biconjugate
gradient linear system solver, written for general use on massively
parallel computers, and our techniques for parallelizing both the
radiation moment solution and the transfer solution.  Additionally,
we present a suite of test problems with a much broader scope than
that covered in the Stone work; new tests include nonequilibrium
Marshak waves, two dimensional ``shadow'' tests showing the one-sided
illumination of an opaque cloud, and full 
RHD+VTEF calculations of radiating shocks.  We use the results of
these tests to assess the virtues and vices of the method as currently
implemented, and we identify a key area in which the method may be
improved.  We conclude that radiation moment solutions closed with
variable tensor Eddington factors show a dramatic qualitative 
improvement over results obtained with flux-limited diffusion, and
further that this approach has a bright future in the context of
parallel RHD simulations in astrophysics.

\end{abstract}


\section{Introduction}

This paper is a logical successor to Paper III of the~\ztwd~series
published in 1992 (see \citet{sto92a}, \citet{sto92b}, and
\citet{sto92c}), which describe numerical methods for carrying out 
radiation magnetohydrodynamic (RMHD) simulations in two dimensions.
In the decade which has passed since these papers
appeared, both the maximum floating point operation (FLOP) speed
and available disk storage capacity have increased by three orders of
magnitude: from gigaflops (GFLOP) to teraflops (TFLOP) in speed;
from Gbytes to Tbytes in storage.  The increase in computing speed
has risen largely from the emergence of massively parallel computer
architectures as the high-performance computing paradigm.  Issues
of cache memory management on modern RISC-based chips, along with
the necessity of writing code for parallel execution, place demands
upon application codes that were largely unknown to the bulk of the
astrophysics community ten years ago.

The spectacular increase in computing power has spawned a new generation
of applications featuring improvements in three major areas: 
higher dimensionality, higher resolution through larger (or sometimes
adaptive) grids, and more realistic physics. 
Early universe simulations and studies of core-collapse supernovae
are but two areas in which increased computing power have profoundly
advanced the realism achievable in a numerical simulation; the lessons
learned in the development of application codes for such problems
are widely applicable to problems throughout astrophysics and
engineering.

As in Paper III of the~\ztwd~series, the focus of this paper is
radiation transport and radiation hydrodynamics; 
in particular, we consider methods which
in principle reproduce and preserve large angular anisotropies in
the radiation field and which treat time dependence in the radiation
field in an appropriate way.  Time dependence and angular anisotropy
highlight two great shortcomings of traditional flux-limited diffusion
(FLD) techniques; the ways in which our method improves upon the
results of FLD form the defining theme of this paper (see Mihalas
and Mihalas (1984) for a good discussion of FLD). The context
of our paper is broader than that of Paper III, however, in that
we have developed new algorithms for simulations on parallel computing
platforms, and we identify key issues which must be addressed for
the successful implementation of a parallel radiation hydrodynamics
(RHD) code.  Additionally,
we present a much more extensive suite of test problems than that
provided in Paper III; of particular interest are the ``shadowing''
tests which, perhaps more than any other, highlight the qualitative
differences between our approach and FLD. 

The impetus for this project was provided by a contract, funded by
the Lawrence Livermore National Laboratory, which supported two of
us (Hayes and Norman) to develop radiative transfer techniques
capable of treating extreme spatial and angular anisotropies in the
radiation field within a medium in which both light-crossing timescales
and (far longer) thermal timescales are important.  The test problem
specified for benchmarking a new algorithm was the ``tophat''
(or ``crooked pipe'') test, a description of which is given by
Gentile (2001).  The algorithm
desired was one that could capture the aforementioned features of
the problem at a fraction of the cost of more elaborate Boltzmann
(e.g. S$_{n}$) or Monte Carlo methods.  We felt that a moment-based
approach like that described in Paper III was an ideal candidate for
treating the tophat test, and further that the original serial 
method could be adapted for parallel use.

The final product of this project is a new set of numerical routines
for performing RHD simulations in a parallel environment.  These
routines provide all of the abilities advertised for the serial
routines in Paper III, and they add new functionality with regard
to both physics and numerics.  In addition, these routines have been
implemented within~\zmp, the latest generation of the ZEUS code series.
The initials ``MP'' refer to the ``multipurpose,'' ``multi-physics''
(HD, RHD, MHD, gravity, chemistry), and ``massively parallel'' aspects
of the code design. The basic HD and MHD equations solved in~\zmp~are
identical to those documented in Paper I and Paper II of the
ZEUS trilogy.  The RHD equations in~\zmp~differ somewhat from those
given in Paper III and are documented extensively in this paper.
The rewriting of the ZEUS algorithm for parallel execution, with
attention given to issues of cache optimization and scalability, has
been documented in a refereed conference proceedings available on
the World Wide Web \citep{fied97}.

This paper is organized in the following manner:~\S{\ref{moment}}
presents the analytic and discrete forms of the RHD moment equations
solved in~\zmp.  \S{\ref{transfer}} presents three different algorithms
for computing the variable tensor Eddington factor (VTEF) used to
close the moment equations.  Two of these algorithms include time
dependence in an approximate way, in contrast to the strictly
time-independent algorithm of Paper III.  \S{\ref{linear}} briefly
describes the new linear solver we have written to solve our 
discrete linear systems on parallel processors, and~\S{\ref{parallel}}
describes the main issues bearing on the implementation of the moment
solution and transfer solution algorithms in a parallel environment.
\S{\ref{tests}} provides a suite of test problems which exercise all
the components of the RHD module.  The main body of the paper concludes
with a summary and discussion (\S{\ref{discuss}}); a full listing of
the linear system matrix comprising our discrete solution to the
radiation moment equations is given in the Appendix.

\section{The Numerical RHD Moment Solution}\label{moment}

As in the~\ztwdft~code described in Paper III,
our algorithm solves the O(1) comoving
equations of radiation hydrodynamics on an Eulerian grid.  Our basic
equations differ from those in~\ztwdft~in that an equation for the
total energy is not solved; rather, we use separate equations for
the gas and radiation energy densities.  While not the optimal choice
in traditional stellar interiors environments, where radiation source
and sink terms are opposite in sign and (very nearly) equal in 
magnitude, our formalism is aimed at environments, both astrophysical
and terrestrial, where matter and radiation are typically well out
of equilibrium; for example, the diffuse interstellar medium or
intergalactic medium.
Additionally, the use of separate gas and radiation energy equations
allows us to employ a numerical scheme which, in addition to being
particularly robust in the nonequilibrium regime, affords a linear
system solution which is more economical than that deployed within
\ztwdft.  Further details on this point are provided in 
\S{\ref{matradcoup}}. We therefore write our equations for the radiating
fluid as:

\begin{equation}\label{cont}
{D\rho \over Dt} + \rho\divv \ = \ 0,
\end{equation}
\begin{equation}\label{gmom}
\rho{D{\bf v} \over Dt} \ = \ -\nabla p + {1 \over c}\chif{\bf F},
\end{equation}
\begin{equation}\label{gase}
\rho\DbDt\left({e \over \rho}\right) \ = \ c\ke E - 4\pi\kp B - p\divv,
\end{equation}
\begin{equation}\label{rade}
\rho\DbDt\left({E \over \rho}\right) \ = \ -\divf - \gradvp +
4\pi\kp B - c\ke E,
\end{equation}
\begin{equation}\label{radm}
{\rho \over c^{2}}\DbDt\left({{\bf F} \over \rho}\right) \ = \ -\divp
- {1 \over c}\chif{\bf F}.
\end{equation}

In (\ref{cont})-(\ref{radm}), $\rho$, {\bf v}, $p$, and $e$ are the
gas density, fluid velocity, gas pressure, and gas energy density,
respectively; $E$, $\F$, and $\p$ are the radiation energy density,
flux, and stress.
Flux-mean, Planck-mean, and energy-mean opacities are defined as
follows:
\begin{equation}
\chif \ = \ {1 \over \F} \int_{0}^{\infty} \chi(\nu)\F(\nu)d\nu,\end{equation}
\begin{equation}
\kp \ = \ {1 \over B} \int_{0}^{\infty} \chi(\nu)B(\nu)d\nu,
\end{equation}
\begin{equation}
\ke \ = \ {1 \over E} \int_{0}^{\infty} \chi(\nu)E(\nu)d\nu.
\end{equation}

In all numerical problems we consider in this paper, $\chif$, $\kp$,
and $\ke$ are equal save for the fact that $\kp$ and $\ke$ are
cell-centered quantities (and directly obtainable from $\rho$ and $e$),
whereas $\chif$ is an interface-averaged variable.  The coordinate
direction along which $\chif$ is averaged is determined by the
component of the flux being evaluated.  We have adopted harmonic
averaging for $\chif$:
\begin{equation}
\chif \ = \ {\chi_{l}\chi_{u} \over {\chi_{l} + \chi_{u}}},
\end{equation}
where the ``l'' and ``u'' refer to cell-centered values on either
side of an interface.

Equations (\ref{cont}) through (\ref{radm}) are closed with expressions
for opacity, an equation of state (currently: ideal gas) and the
following radiation variables:
\begin{equation}\label{planck}
B = {caT^{4}\over 4\pi},
\end{equation}
and
\begin{equation}\label{eddtens}
\p \ = \ \f E.
\end{equation}
Equation (\ref{planck}) defines the familiar grey Planck function,
and (\ref{eddtens}) expresses the radiation stress tensor, $\p$,
in terms of the Eddington tensor, $\f$:
\begin{equation}\label{fedd}
\f \ \equiv \ {\p \over E} \ = \ {\oint I{\bf nn} d\Omega \over
                                  \oint I         d\Omega},
\end{equation}
Here $I$ is the space- and angle-dependent specific intensity of
the radiation field, {\bf n} is the local unit normal vector, and
$d\Omega$ is an element of solid angle.

     \subsection{Operator Splitting}

\zmp~employs the same operater splitting scheme for evolving the
RHD equations as that employed in earlier ZEUS codes and documented in
\citet{sto92a}.  In this formalism, the solution is divided into
``source'' and ``transport'' steps.  In the source step, we update
the radiation moment variables according to:
\begin{eqnarray}
{\partial e \over \partial t} & = & c\ke E - 4\pi\kp B - p\divv,
\label{gase2} \\
{\partial E \over \partial t} & = & 4\pi\kp B - c\ke E -\divf
 - \gradvp, \label{rade2} \\
{\partial \F \over \partial t} & = & -c\chif \F - c^{2}\divp;
\label{radm0}
\end{eqnarray}
while in the transport step, we solve:
\begin{eqnarray}
\dbdt\int_{V}e dV & = & -\oint_{dV}e({\bf v} - {\bf v_{g}})\cdot
 d{\bf S}, \\
\dbdt\int_{V}E dV & = & -\oint_{dV}E({\bf v} - {\bf v_{g}})\cdot
 d{\bf S}, \\
\dbdt\int_{V}\F dV & = & -\oint_{dV}\F({\bf v} - {\bf v_{g}})\cdot
 d{\bf S},
\end{eqnarray}
where ${\bf v_{g}}$ is the local grid velocity.

     \subsection{Radiation Energy Density and Flux}

Our numerical prescription for the radiation flux differs slightly
from that presented in Paper III.  In the previous work, the
radiation flux was written according to the ``Automatic Flux-Limiting''
prescription of \citet{mih82}.  In this work, we adopt a standard
fully implicit time differencing form for the flux equation, although
we express it as a function of the updated radiation energy density,
as in the previous work.  To compare, we begin by writing (\ref{rade2}),
making use of (\ref{eddtens}), as
\begin{equation}\label{radm1}
{\partial \F \over \partial t} \ = \ -c\chif \F - c^{2}\divfe.
\end{equation}

In the AFL prescription, (\ref{radm1}) is integrated analytically
over a timestep, $\Delta t$, to yield:
\begin{equation}\label{radm2}
\F\supnp\afl \ = \ e^{-c\chif\Delta t}\F\supn \ - \ 
\left(1 - e^{-c\chif\Delta t}\right){c \over \langle\chif\rangle}
\langle\divfe\rangle,
\end{equation}
where the quantities in $\langle\rangle$ brackets indicate an
average value for the timestep.  Noting, however, that (\ref{radm1})
may be written as a finite time difference:
\begin{eqnarray}
{\F\supnp - \F\supn \over \Delta t} & = & -c\theta\chif\supnp\F\supnp
-c^{2}\theta\deldot(\f E\supnp) \nonumber \\
 & - & c(1 - \theta)\chif\supn\F\supn
 -c^{2}(1- \theta)\deldot(\f E\supn),
\end{eqnarray}
we may write an analogous ``implicit time difference'' form of
(\ref{radm1}) as
\begin{equation}\label{itdflux}
\F\supnp\itd \ = \ {\F\supn \over {1 + c\chif\supnp\Delta t}} - 
{{c^{2}\Delta t}
\over {1 + c\chif\supnp\Delta t}}\deldot(\f E\supnp).
\end{equation}

Writing $x \equiv c\chif\Delta t$, $\F\supnp\afl$ and $\F\supnp\itd$
are expressible as:
\begin{eqnarray}
\F\supnp\afl & = & e^{-x}\F\supn - \left(1 - e^{-x}\right)
\left(c \over \chif\supnp\right)\deldot(\f E\supnp); \\
\F\supnp\itd & = & {\F\supn \over {1 + x}} - {x \over {1 + x}}
{c \over \chif\supnp}\deldot(\f E\supnp).
\end{eqnarray}
Here, we have written $\F\supnp\afl$ for the case where the bracketed
quantities are represented by their values at the advanced time.
In the limit that $x$ approaches zero (small timesteps and/or
transparent media, one may show that
\begin{eqnarray}
\F\supnp\afl & \rightarrow & (1 - x)\F\supn - x
\left(c \over \chif\supnp\right)\deldot(\f E\supnp); \\
\F\supnp\itd & \rightarrow & (1 - x)\F\supn - x
{c \over \chif\supnp}\deldot(\f E\supnp).
\end{eqnarray}
Similarly, when $x$ is very large, both expressions achieve a
form resembling a diffusion equation:
\begin{eqnarray}
\F\supnp\afl & \rightarrow & - \left(c \over \chif\supnp\right)
\deldot(\f E\supnp); \\
\F\supnp\itd & \rightarrow & - \left(c \over \chif\supnp\right)
\deldot(\f E\supnp).
\end{eqnarray}

In the course of testing our code, we tried both methods when
running test problems such as those shown in section~\ref{tests}.
We found no real practical difference between the two approaches,
and have chosen $\F\supnp\itd$ to represent the flux at the advanced
time.  The virtue of both approaches is that $\F\supnp$ is written
as a function of the updated $E\supnp$ and can therefore be
substituted algebraically into (\ref{rade}), thus reducing our
system of independent equations (and variables) by one.  When
used in conjunction with the operator splitting scheme for
matter-radiation coupling (next section), it then becomes necessary
to construct and solve a linear system matrix for only one unknown
moment variable, in contrast with the approach employed in \ztwdft.

     \subsection{Matter-Radiation Coupling}\label{matradcoup}

A major difference between our treatment of the radiating fluid
and that used in \ztwdft~lies in our use of a separate equation
for the gas energy density alone, as opposed to the total energy
equation used in \ztwdft.  Furthermore, we have chosen an operator
splitting scheme for matter-radiation coupling which requires a
linear system for only the radiation energy density to be solved.
The scheme in \ztwdft~is particularly attractive in regimes where
$\F \ll cE$ and the matter and radiation are nearly in equilibrium;
the total energy equation does not depend on radiation source/sink
terms which are extremely large and of opposite sign.  The cost of
solving the linear system is higher, however, for one must solve two
coupled equations in two unknowns at each mesh point.  The scheme
described here is a frequency-integrated version of a multi-frequency
method documented by \citet{bal99}.  It has the twin virtues of 
being particularly robust for problems where radiation
and matter are far out of equilibrium, and of allowing a linear system
in $E$ alone to be constructed.

Making use of our formula for the flux (equation~\ref{itdflux}), we
may rewrite (\ref{rade2}) as,
\begin{equation}\label{rade3}
E\supnp - E\supn \ = \ \Delta t \{ 4\pi\kp B - c\ke E\supnp
 - {\cal{G}}_{1}(E\supnp) - {\cal{G}}_{2}(\F\supn)
 - {\cal{H}}(E\supnp) \},
\end{equation}
where we have defined ${\cal{G}}_{1}$, ${\cal{G}}_{2}$, and
${\cal{H}}$ as
\begin{equation}\label{g1term}
{\cal{G}}_{1} \ \equiv \ -\deldot
\left\{\left(x \over 1 + x \right)\left({c\over\chif}\right)
\divfe\supnp\right\},
\end{equation}
\begin{equation}
{\cal{G}}_{2} \ \equiv \ \deldot\left\{\left(1 \over 1 + x\right)
\F\supn\right\},
\end{equation}
and
\begin{equation}
{\cal{H}} \equiv (\nabla{\bf v} : \f) E\supnp.
\end{equation}
 
Our gas energy equation, ignoring the $PdV$ work term, is
\begin{equation}\label{gase3}
e\supnp - e\supn \ = \ \Delta t \{ c\ke E\supnp - 4\pi\kp B \}.
\end{equation}
The contribution from the work term is performed at a later stage in
the ``source step'' update.  The heart of the coupling scheme is
the linearization of the Planck source function through a Taylor
series expansion:
\begin{equation}\label{plancklin}
B \ \simeq \ B(T\supn) \ + \ \left(T\supnp - T\supn\right)
{dB \over dT}\Bigg|_{T^{n}}.
\end{equation}
This allows us to approximate the gas temperature at the advanced
time in terms of the old temperature as
\begin{equation}\label{lint}
(T\supnp)^{4} \ \simeq \ \left(T\supn\right)^{3}\left\{4T\supnp - 
3T\supn\right\}.
\end{equation}

To proceed, we express the gas energy in terms of the specific
heat at constant volume and the gas temperature:
\begin{equation}\label{egas}
e\supnp     \ \equiv \ \rho c_{v}T\supnp,
\end{equation}
and we define the following coupling coefficients and coupling
function:
\begin{eqnarray}
k_{1} & \equiv & c\kp\Delta t; \\
k_{2} & \equiv & c\ke\Delta t; \\
\Phi & \equiv & 1 \ + \ \left(4ak_{1}\over \rho c_{v}\right)
\left(T\supn\right)^{3}\label{phifunc}.
\end{eqnarray}
We note that the opacities are assumed to be evaluated at the old
material temperature and thus known.  Armed with expressions
(\ref{egas}) through (\ref{phifunc}), it is straightforward to
transform (\ref{gase3}) into an equation for the updated material
temperature:
\begin{equation}\label{tgas}
T\supnp \ = \ \left(1\over\Phi\right)\left\{T\supn \ + \
\left(1\over\rho c_{v}\right)\left[3ak_{1}\left(T\supn\right)^{4}
 \ + \ k_{2}E\supnp\right]\right\}.
\end{equation}
This expression for $T\supnp$ is used to evaluate $B$ according
to (\ref{plancklin}), and is then substituted into (\ref{rade3}).
Performing the algebra and grouping terms appropriately yields
the following expression for the radiation energy density at the
advanced time:
\begin{equation}\label{rade4}
\left\{1 \ + \ {k_{2}\over\Phi}\right\}E\supnp \ + \ \Delta t
({\cal{G}}_{1}(E\supnp) + {\cal{H}}(E\supnp)) \ = \ E\supn \ - \
\Delta t{\cal{G}}_{2}(\F\supn) \ + \ 
{k_{1}a\left(T\supn\right)^{4} \over\Phi}
\end{equation}

Equation (\ref{rade4}) expresses $E\supnp$ in terms of known
opacities and quantities from the previous timestep: $E\supn$,
$T\supn$, and $\F\supn$.  Furthermore, the update of the radiation
flux is built in through the expression for ${\cal{G}}$.  When
the divergence operators acting on $\F$ and $\f E$ are written
out in finite difference form (see Appendix A for details), a
matrix equation results which couple values of $E\supnp$ on a
nine-point stencil:
\begin{eqnarray}\label{radmat}
\begin{array}{lclclc}
{\rm DM4}\subij \ E\supnp\subimjm & + &  
{\rm DM3}\subij \ E\supnp\subijm  & + &  
{\rm DM2}\subij \ E\supnp\subipjm & +  \nonumber \\
{\rm DM1}\subij \ E\supnp\subimj  & + & 
{\rm D00}\subij \ E\supnp\subij   & + &  
{\rm DP1}\subij \ E\supnp\subipj  & + \nonumber \\
{\rm DP2}\subij \ E\supnp\subimjp & + & 
{\rm DP3}\subij \ E\supn\subijp   & + & 
{\rm DP4}\subij \ E\supnp\subipjp &   
\end{array} 
\nonumber \\ 
\nonumber \\ \ = \ {\rm RHS}(E\supn,T\supn,\F\supn)\subij
\end{eqnarray}
In (\ref{radmat}), $D00$ represents elements along the main diagonal
of the matrix, which multiply solution vector ($E$) elements at
the central point of the finite-difference stencil.  Similarly,
$DM1$...$DM4$ represent subdiagonal bands in the matrix, and
$DP1$...$DP4$ indicate the superdiagonal bands.

Once (\ref{radmat}) is evaluated, the new temperature is computed
according to (\ref{tgas}) and transformed into the gas energy through
(\ref{egas}).  Additionally, the fluxes are updated via (\ref{itdflux})
with the new values of $E\supnp$.  With new values of $e$ and $E$,
the Eddington tensor may be updated if necessary.  We note that a
fully self-consistent mathematical treatment would involve a grand
iterative solution whereby $E$, $e$, and $\f$ are iterated to
convergence.  In reality, such a procedure is expensive even in one
dimension, and prohibitive in multidimensional calculations.  We
therefore regard $\f$ as known once it is computed with updated
source and sink terms.  Furthermore, we allow the code to proceed
with the same values of $\f$ for multiple timesteps when the 
matter temperature is evolving slowly.  We monitor the maximum 
cumulative fractional change in $T$ over the entire computational
grid and update $\f$ when, at some location, $\Delta T/T$ has changed
by some preset tolerance since the last $\f$ update.  The typical
value of this tolerance is on the range of 2-5\%.

The moment equations described above are coupled to the gas dynamical
equations and solved at each timestep.  A full solution to the RHD
equations may be completed if $\f$ is known; our three techniques
for computing it are detailed in~\S{\ref{transfer}}.

\subsection{Discretization}

We conclude this section by noting that~\zmp~employs the same
discretization techniques used in previous ZEUS codes.  Equations
are differenced using a staggered mesh defining both cell-centered
variables ($\rho$, $e$, $E$, $p$, opacities, and diagonal elements
of $\p$), interface variables ({\bf v} and $\F$), and corner variables
(off-diagonal elements of $\p$).  In addition, the discrete equations
make use of the covariant metric coefficients documented in
\citet{sto92a}; these coefficients appear extensively in the
documentation (see Appendix~\ref{matdoc}) of the linear system matrix
generated from our discrete radiation moment equations.

\section{Closing the Moment Equations: The Transfer Solution}
\label{transfer}

Paper III describes a scheme for solving
the static (time-independent) transfer equation, in two-dimensional
axisymmetric geometry, by the method of short characteristics (SC)
on tangent planes.  This approach forms the foundation of our method,
and we have added considerable functionality, with respect to both
physics and numerics, on top of the original scheme.  Because Stone's
algorithm and ours share a common origin, we will not reproduce the
full discussion of the tangent-plane method and the static SC solution
provided in the 1992 paper.  We note, however, that the calculation
of quadrature weights for performing $\mu$- and $\Phi$-angle integration
of $I$ is identical in both codes (cf. eq. 75-89 of Paper III).  
The machinery for assembling
angle-averaged moments of $I$ is also identical, save for the ability
of our code to parallelize (to some degree) the integration over
$\mu$.  Finally, the coefficients used to interpolate $I$ along
grid faces in a tangent plane (eq. 58-62 of Paper III) are
identical.

To proceed, we will document the
ways in which our new algorithm differs from and enlarges upon the
machinery in \ztwdft.  Modifications with respect to physics
involve the addition, in two different approximations, of time
dependence to the formal solution.  While our code retains the 
static formal solution as an option, we focus in this paper
primarily upon transport solutions
that include time dependence, descriptions of
which follow in this section.  Modifications with
regard to numerics involve two components: (1) the addition of
computational parallelism to the SC framework, and (2) the option
of solving the transfer equation on a ``coarse grid'' sampled from
a higher-resolution ``fine grid'' on which the RHD equations are solved.
In later discussions we will refer to this approach as the ``coarsened
short characteristics'' (CSC) method.  This option, in combination
with parallel execution, is of high value when a time-dependent
transfer solution is desired.  As \citet{sto92c} point out, a static
transfer solution does not require the specific intensity to be
stored for any points other than those on the tangent plane currently
being treated.  Therefore, a two-dimensional array of size equal to
that needed for the moment solution field variables
provides sufficient storage for $I$ at any given instant.
By contrast, time-dependent treatments require a complete
specification of $I$ from a previous timestep, which means that the
full run of $I$ with space and angle must be stored.  This requirement
mandates a four-dimensional array for $I$ in a spatially two-dimensional
problem.  
Time dependence therefore introduces a large memory burden which
rapidly becomes prohibitive as grid size increases.  Our parallel
CSC approach, however, relieves this burden to a considerable degree
and enables RHD simulations with
time-dependent VTEF solutions on much larger grids than otherwise
attainable.  The CSC approach for Eddington tensors is
discussed at the end of this section; a discussion of parallelism
within the formal solution appears in section~\ref{parallel}.

 \subsection{The TRET Algorithm: Time-Retarded Opacities}

The SC solution for Eddington tensors is rooted in a solution to
the transfer equation along a ray of arbitrary orientation in space:
\begin{equation}
{1 \over c}{\partial I \over \partial t} \ + \ 
{\partial I \over \partial l} \ = \ \eta - \chi I,
\label{teq}
\end{equation}
where $l$ denotes a ray along which $I$ is to be computed.
The {\it formal solution}, as traditionally defined, to (\ref{teq})
results when (1) the time derivative term is dropped, and (2) the source
function ($\eta / \chi$) is assumed known {\it a priori}.  Writing
the source function as $S$ and transforming from spatial to optical
depth coordinates, we obtain the familiar static form of the transfer
equation:
\begin{equation}
{\partial I \over \partial \tau} \ = \ I - S,
\label{stateq}
\end{equation}
which has the following solution:
\begin{equation}
I_{d} \ = \ I_{u}e^{-\tau_{L}} \ + \ \int_{0}^{\tau_{L}}
S(\tau)e^{-\tau}d\tau,
\label{statsol}
\end{equation}
where $I_{u}$ and $I_{d}$ are the specific intensities at the upstream
and downstream points, respectively, on a ray segment of length $L$.
The algorithm for computing Eddington tensors described in 
Paper III is rooted in (\ref{statsol}).  This is
a static solution to the transfer equation, and is therefore appropriate
only when timescales of interest are much longer than radiation flow
timescales.  When this condition is not met, then some means of
including time dependence when evaluating $I$ is needed.  In this
section we describe the first of two methods we have implemented
toward this end.  In common with the original static approach, we
drop the time derivative from (\ref{teq}).  In contrast with the static
approach, however, we evaluate the source function at the appropriate
retarded time when integrating along a characteristic ray.  In the
Stone algorithm, the source function is allowed to vary spatially
over the grid cell spanned by a characteristic ray, but all quantities
in (\ref{statsol}) are evaluated at the advanced time.  In the 
time-retarded approach, material properties are assumed to be spatially
uniform over the cell, but are additionally assumed to vary in time. 
Integration over an optical path length, such as that indicated
in (\ref{statsol}), then involves in implied integration over
an appropriate interval of retarded time:
\begin{eqnarray}\label{treteq}
I_{d}(t_{L}) & = & 
 I_{u}exp{\left[-\int_{0}^{L}\chi_{c}\left(t_{L}-{l \over c}\right)dl 
\right]} \nonumber \\
 & + & \int_{0}^{L}S_{c}(t_{L}-{l \over c})
 exp{\left[-\int_{0}^{L}\chi_{c}\left(t_{L}-{l' \over c}\right)dl' \right]}
 \chi_{c}\left(t_{L}-{l \over c}\right)dl.
\end{eqnarray}
Here, $t_{L}$ is the time required for a photon to traverse a distance
$L$ from the upstream point to the downstream point, at which the
solution is desired.  The arguments for $S_{c}$ and $\chi_{c}$ are
the proper retarded times for the spatial coordinate, $l$.
The subscript ``c'' on the source function and
opacity indicate that these are cell-averaged values.

Equation~\ref{treteq} is a special case of a more general form discussed
in~\citet{mih84}. The method by which we evaluate the formal solution
along characteristic rays is taken from a treatment outlined in
~\citet{mih99}.  Because material properties are assumed to be
spatially uniform across a cell, the integrals over optical depth
are straightforward.  Begin by defining:
\begin{equation}
\chi_{u} \ \equiv \ (1-\theta)\chi_{0} \ + \ \chi_{-1},
\end{equation}
and
\begin{equation}
S_{u} \ \equiv \ (1-\theta)S_{0} \ + \ S_{-1},
\end{equation}
where the subscript ``u'' denotes quantities at the upstream point
at the advanced time, ``0'' denotes cell-centered quantities at
the advanced time, and ``-1'' denotes cell-centered quantities at
the previous time.  The $\theta$ parameter is a time interpolant
defined as follows:
\begin{equation}\label{thetapar}
\theta \ \equiv \ {\rm min}\left(1,{L \over c\Delta t^{*}}\right).
\end{equation}
Note that the timestep value, $\Delta t^{*}$, is labeled with a
star to indicate that it is the elapsed simulation time between
successive updates of $I$ and need not in general equal the
hydrodynamic timestep, $\Delta t$.
Since the speed of light is constant, the interpolation for
$\chi$ and $S$ is equivalent to the following:
\begin{equation}
\chi(l) \ = \ \chi_{0} + \left(\chi_{u}-\chi_{0}\right)\left(
{l \over L}\right),
\end{equation}
and
\begin{equation}
S(l) \ = \ S_{0} + \left(S_{u}-S_{0}\right)\left(
{l \over L}\right).
\end{equation}
The optical depth along a ray is then given by:
\begin{equation}\label{tauvsl}
\tau(l) \ = \ \int_{0}^{l}\chi(l')dl' \ = \ \chi_{0}l +
\left({\chi_{u} - \chi_{0}} \over {2L}\right)l^{2}.
\end{equation}

With opacities and source functions defined as described, the
time-retarded solution for the specific intensity is written
as
\begin{equation}
I_{d} \ = \ I_{u}e^{-\tau_{L}} \ + \ 
\int_{0}^{\tau_{L}} \left[S_{0} + \left({S_{u}-S_{0}}\over L\right)
l(\tau)\right] e^{-\tau}d\tau.
\label{treteq2}
\end{equation}
Given an appropriate value of the upstream intensity, $I_{u}$,
a solution to (\ref{treteq2}) is obtainable if we can express
the integral path length, $l$, as a function of the optical depth.
In the simple test problems for which this method was used, the
opacities were independent of temperature.  In this case, we
have that $\chi_{u} = \chi_{0}$, and (\ref{tauvsl}) reduces to
\begin{equation}
\tau(l) \ = \ \chi_{0}l.
\end{equation}
Equation (\ref{treteq2}) then has the following solution:
\begin{equation}\label{tretsol}
I_{d} \ = \ I_{u}e^{-\tau_{L}} \ + \ S_{0} \ - \ S_{u}e^{-\tau_{L}}
 \ + \ \left(S_{u} - S_{0} \over \tau_{L}\right)\left(
 1 - e^{-\tau_{L}}\right)
\end{equation}

$I_{u}$  must be temporally interpolated from values at the advanced
and previous times, so we write
\begin{equation}\label{iselect}
I_{u} \ \equiv \ (1 - \theta)I_{u}^{n+1} + \theta I_{u}^{n}.
\end{equation}
Here, ``n'' and ``n+1'' refer to the previous and advanced times,
respectively.

In future reference, our algorithm based upon (\ref{treteq2}) will
be called the TRET method.
There are two features of this approach which merit comment.
For the case of constant opacities,
equation (\ref{tretsol}) is identical in form to equation (69) of
Paper III.  This means that the SC machinery developed for
the static solution can serve as a template for the TRET code
in the case of temporally constant opacity.  Numerically, the most significant
difference between the two codes is memory requirement: not only
must opacities and emissivities from a previous timestep be stored
for a TRET solution, but the full four-dimensional specific intensity
must be saved as well.  This latter requirement makes the TRET
algorithm, as well the PSTAT code described in the next section,
far more memory intensive than the purely static solution.  The
large memory requirements imposed by the time-dependent methods have
led to the development of an angle-parallel SC method, which is
discussed in \S~\ref{parallel}.

The second feature concerns the degree to which time-dependent
variations in $I$ can be accurately represented.  Note from
(\ref{thetapar}) that when $\Delta t^{*} > L/c$, where $L$ is the
length of a characteristic, the value of $I$ at the upstream boundary
of the cell will be taken at the advanced time.  Considering rays
which lie along the Z-axis in cylindrical geometry ($\mu$ = 0),
the condition $\Delta t^{*} > L/c$ will be simultaneously satisfied for
{\it all cells} in the case of a uniform grid.  In this instance,
all cells along the full ray will be spatially coupled, and the
only variations of $I$ will be due to sources and sinks along the
ray.  Considering the case of a plane wave propagating through vacuum
(an effectively one-dimensional problem), we see that changes in
$I$ at an illuminating source boundary will be propagated across the
entire domain, even though time retardation is present in the solution.
Thus this method does not possess the same degree of temporal accuracy
that a solution to the fully time-dependent transfer equation should
have.  An exception to this behavior occurs, however, when 
$\Delta t^{*}$ is exactly equal to $L/c$.  If there are no intervening
sources or sinks of radiation (i.e. a vacuum) along the ray, then the
use of (\ref{iselect}) in (\ref{tretsol}) will cause $I_{d}$ to equal
$I_{u}$.  Consider a vacuum environment in which the radiation field
is initially uniform at a value $I_{o}$, and imagine that an
illuminating source, $I_{s}$, is initialized at the domain boundary
on one end at t = 0.  Under the conditions we have identified, the
source value $I_{s}$ will propagate one cell width at every timestep.
Assuming the timestep is $L/c$, it is possible to force the 
algorithm into propagating a sharp wavefront along a one-dimensional grid
causally.  We emphasize, however, that this is a special case and
not generally applicable to problems of interest.

 \subsection{The PSTAT Algorithm: An Approximate Time Derivative Operator}

Our second time-dependent method retains the time derivative in
(\ref{teq}), albeit in an approximate way.  We replace the analytic
partial derivative with a finite time difference:
\begin{equation}
{1 \over c}{\partial I \over \partial t} \ \rightarrow \ 
{{I - I^{n}} \over c\Delta t^{*}}.
\end{equation}
Here, $I$ is the intensity at the desired advanced time, and $I^{n}$
is a previous solution separated in time from $I$ by $\Delta t^{*}$.
Recall that $\Delta t^{*}$ in general 
need not equal the hydrodynamic timestep, $\Delta t$.  If the 
radiation field is evolving slowly, then the Eddington tensors
may not require an update at every timestep.  In this case $\Delta t^{*}$
becomes the time between successive calls to the transfer solution
algorithm.  With the approximate time derivative defined in this
way, (\ref{teq}) becomes
\begin{equation}
{\partial I \over \partial l} \ = \ \tilde{S} - \tilde{\chi} I,
\label{pstat}
\end{equation}
where
\begin{equation}
\tilde{S} \ \equiv \ \chi S \ + \ {I^{n} \over c\Delta t^{*}},
\end{equation}
and
\begin{equation}
\tilde{\chi} \ \equiv \ \chi \ + \ {1 \over c\Delta t^{*}}.
\end{equation}

Because (\ref{pstat}) functionally resembles its static analog in
(\ref{stateq}), we refer to this approach as the {\it
pseudostatic} solution to the transfer equation, and the computer
algorithm based upon this solution is denoted with the PSTAT label.
Considering the
propagation of radiation along a ray of length $L$, and assuming
that $S$ and $\chi$ are uniform along the ray (appropriate for a grid
cell) the solution to (\ref{pstat}) follows immediately:
\begin{equation}
I_{d} \ = \ I_{u} e^{-L\tilde{\chi}} \ + \ {\tilde{S}\over\tilde{\chi}}
\left(1 \ - \ e^{-L\tilde{\chi}} \right).
\label{pstatsol}
\end{equation}
In (\ref{pstatsol}), $I_{d}$ and $I_{u}$ have the same meanings as
in (\ref{statsol}).
In the limit of very large timesteps, $\tilde{\chi}$ approaches $\chi$
and $\tilde{S}/\tilde{\chi}$ approaches $S$, so that the proper static
limit is recovered.  As $\Delta t^{*}$ approaches zero, the exponential
terms in (\ref{pstatsol}) vanish, and $\tilde{S}/\tilde{\chi}$
becomes $I^{n}$, as required.

Unlike the TRET algorithm, the PSTAT code does not require that
$S$ and $\chi$ from a previous timestep be stored, which results
in some memory savings.  The PSTAT approach is also simpler to
implement than the TRET method, and it has been widely used
in terrestrial transport applications (e.g. Adams 1997). We have
therefore adopted it as our default method for extracting time-dependent
transport solutions.

As with the purely static and TRET solutions, the solution
to $I$ from the PSTAT algorithm involves a summation of
terms which exponentially decay over space.  In general, then, it is
not possible to reproduce pure plane-wave ``step function'' profiles
in the radiation energy density, save for a special case in the
TRET approach which we mentioned earlier.

     \subsection{Eddington Tensors on Large Grids: Coarsened Short
                 Characteristics}

We have noted previously that SC solutions to the transfer equation
which include time dependence are extremely memory intensive.  This
is due to a combination of factors.  The first is that time-dependent
solutions require that the specific intensity at a given update
be stored for use at the next update.  In contrast with the static
algorithm, where $I$ only need be saved on a single tangent plane
for a single $\mu$ angle cosine, $I$ must be saved for all spatial
and angular points.  This mandates four-dimensional array storage
for a problem in two spatial dimensions.  The second factor involves
the selection of the angles themselves.  Within the tangent-plane
method, the user is free to determine the number of $\mu$ angle cosines
used, independent of the number of axial and radial zones.  The
$\Phi$ angles, however, result from the intersections of tangent
planes with the cylindrical shells defined by the array of radial mesh
points.  The number of $\Phi$ angles are thus fixed by the size
of the moment solution grid.  The total memory needed therefore scales
as the square of the number of radial mesh points, but is only linear
in the number of axial mesh points and $\mu$ angular rays.

The relationship of the $\Phi$ grid to the moment grid makes 
time-dependent VTEF calculations impossible on current CPU
architectures once the number of radial mesh points exceeds values
of order 100, unless very small axial or $\mu$ angle meshes are
used.  To enable calculations with good resolution along both
coordinate axes, we have developed a {\it Coarsened Short
Characteristics} (CSC) approach, in which the Eddington tensors are
computed on a grid at lower resolution with respect to the moment
solution grid.  Our approach is conceptually quite simple.  Given
a ``fine grid'' upon which the gas dynamic and radiation moment
equations are evolved, we may extract a ``coarse grid'' for evaluating
Eddington tensors by sampling the fine grid at a regular interval
along each coordinate axis.  The number of fine-grid and coarse-grid
points, $N_{\rm fg}$ and $N_{\rm cg}$, are related to the sampling
frequency, $N_{\rm samp}$ by
\begin{equation}
N_{\rm fg} \ = \ N_{\rm samp}\cdot\left(N_{\rm cg} - 1\right) \ + \ 1
\label{sampform}
\end{equation}

\begin{figure}
\plotone{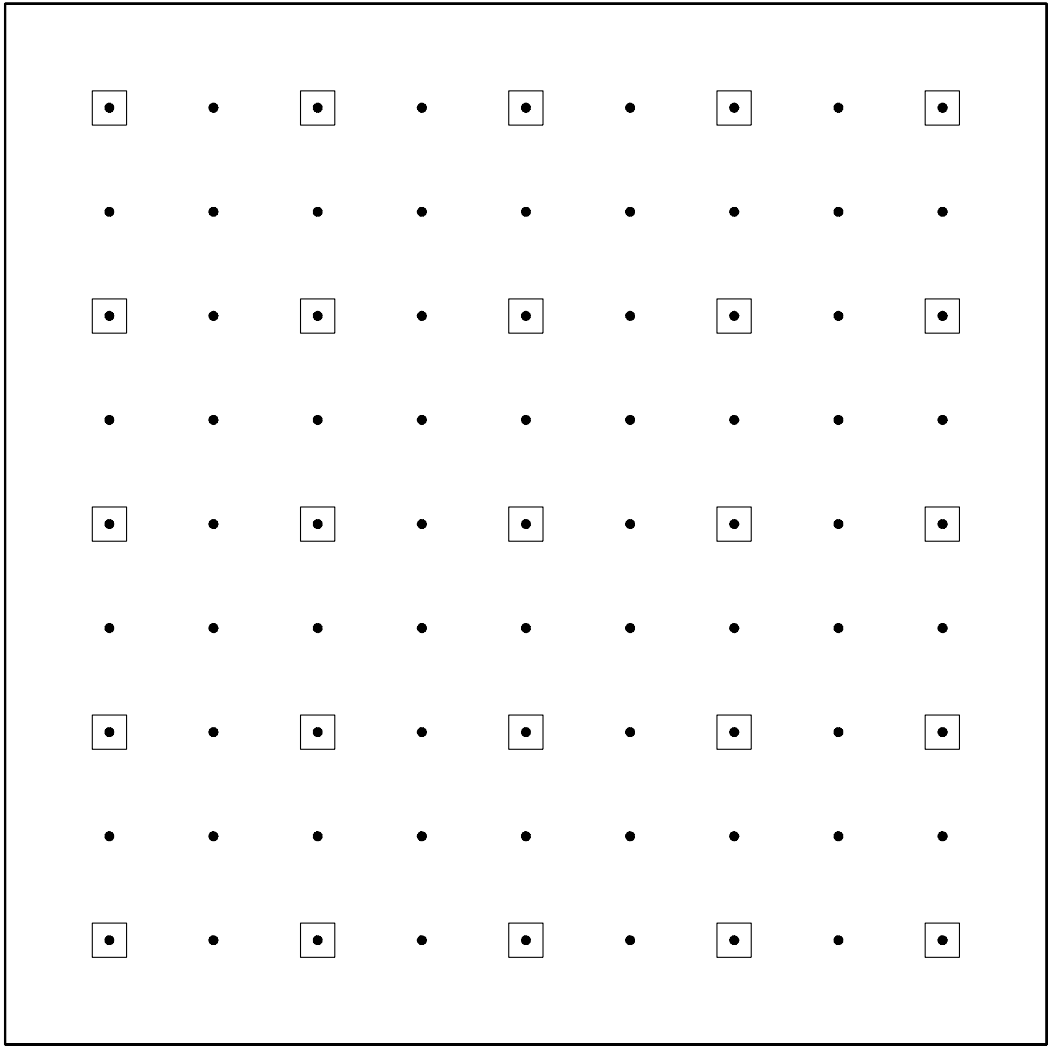}
\caption{The distribution of coarse-grid points, used for the CSC
transfer solution, relative to the full-resolution moment solution
grid.  The case of $N_{\rm samp}$ = 2 is shown.  The fine grid is
shown by the small dots; the open squares indicate the mesh points
selected for the transfer solution.  In the case of a spatially
parallelized calculation, this pattern is replicated on the local
subgrid contained on each processor.\label{fig1}}
\end{figure}

Figure~\ref{fig1} illustrates the relative placement of the fine
and coarse grids for a 9x9 fine grid sampled with a frequency of
2.  Note that the boundary values of the coarse grid coincide
with boundary values of the fine grid.  This is deliberately enforced
to facilitate parallel calculations, when local subgrid boundary data
must be exchanged between processors.  In the case of a parallel
calculation, the arrangement in figure~\ref{fig1} would represent the
fine and coarse grids on the local subdomain owned by a particular
CPU and stored in that processor's memory.  If one coordinate axis
of the problem is divided into $N_{\rm proc}$ subdomains and
distributed to $N_{\rm proc}$ CPU's, then the global fine and
coarse grids are equal to $N_{\rm proc}$ times the local portions
appearing in (\ref{sampform}).  This arrangement can be applied
independently along each axis, and $N_{\rm proc}$ need not be the
same for both (although currently $N_{\rm samp}$ is the same in
both directions). We therefore see that the numbers of {\it global} fine
and coarse grid points do not follow the relationship in 
(\ref{sampform}) in a parallel calculation.
The exact relationship depends upon the parallel topology of the
calculation.

Given that we ultimately require knowledge of the Eddington tensor,
$\f$, on the global
fine grid, the CSC solution for $\f$ requires a subsequent interpolation
step.  For our initial development, we have used bilinear
interpolation to construct fine-grid Eddington tensors from the
coarse-grid solution.  Bilinear interpolation was chosen for its
obvious simplicity and ease of implementation, and has proven adequate
to the task of demonstrating potential viability of the CSC approach.
Nonetheless, we acknowledge that such a low-order scheme is likely
to introduce errors that may be avoided by more sophisticated methods.
Such an investigation was not deemed necessary for this introductory
report, and will be pursued, if needed, in future work.

\section{The Linear System Solver}\label{linear}

The implicit solution for the radiation field variables necessitates
an efficient algorithm for solving a large, sparse linear system.
The requirement of efficient parallel execution on a large number
of CPU's places further constraints on the potential menu of solution
methods.  Krylov methods
are known to have particular utility in this context, and we have
chosen to develop a parallel linear solver package based upon the
pre-conditioned {\it biconjugate gradient} method (BiCG).
Pseudocode templates for the BiCG algorithm and other methods in
the Krylov subspace family are available in \citet{bar94}.

Because the BiCG technique is an iterative method, its computational
performance hinges upon how rapidly convergence is
reached.  A fact sometimes overlooked in discussions of the
performance of various linear solution techniques is that the
performance of a given technique is often extremely problem
dependent.  Furthermore, the convergence rate of a linear system
solver can vary widely over the course of a time-dependent simulation.
Why this may happen in an astrophysical simulation may be understood
simply as follows: the matrix represented symbolically by equation
(\ref{radmat}) possesses elements along the main diagonal with
the following form: ${\rm D00}\subij \ = \ 1 - \Delta t \times
({\rm terms})$, where
$\Delta t$ is the timestep.  Elements along the subdiagonal and
superdiagonal bands
are directly proportional to $\Delta t$.  Thus when $\Delta t$ is
small, the matrix becomes increasingly diagonally dominant, and tends
toward the identity matrix in the limit of small timesteps.
Since the size of $\Delta t$ is regulated by a number of highly
variable constraints (e.g. fractional changes in field variables,
the Courant time, etc.), the diagonal dominance of the matrix may
change strongly during a simulation.  As noted by \citet{bal99},
this behavior can be so strong that very different iterative methods
become optimal during different stages of a simulation.  This suggests
that adaptive switching between solution methods can be a profitable
feature of an application code, but such effort is beyond the scope
of this current project.

\section{Implementing Parallelism}\label{parallel}

From the standpoint of performance, the design of this code for
parallel execution represents the greatest change from the algorithms
written for \ztwdft.  The desire to model phenomena in multidimensions
at high spatial resolution places severe demands both on needed
memory storage and the number of floating point operations required
for a numerical solution.  In addition to simply requiring that more
grid points be included in a simulation, astrophysicists are faced
with the reality that many phenomena of interest involve a large
number of complex physical processes acting in concert, such as
radiation transport, general relativity, nuclear burning,
multidimensional fluid flow, and multispecies mixing and transport.
Arrays of parallel processors, rather than
more powerful single-CPU machines, have claimed the top end of the
high-performance computing domain; researchers who model complex
astrophysical systems (e.g. Type I and Type II supernova explosions,
large-structure formation in the early universe) are thus finding
it necessary to become literate in issues once the (almost) exclusive
province of computer scientists.

Adopting the terminology set forth in \citet{fos95}, we identify
{\it domain decomposition} as our model for parallel execution.
In this model, we imagine the
computational domain as being physically divided into separate
portions, each of which is operated upon by a unique computer
processor.  Each processor performs an identical set of operations
upon its share of the data. For the majority of the subroutines
in~\zmp, the data is spatially decomposed; i.e. each processor
contains a physically contiguous subset of the
computational grid.  This approach forms the basis for parallel
execution of all subroutines concerned with hydrodynamic evolution
and the radiation moment solution.
The calculation of Eddington tensors, however, requires
a different approach.  Characteristic solutions to the transfer
equation are by nature spatially recursive, thus the computation
of the specific intensity along a given angular ray is intrinsically
serial: the calculation at one point may not begin until that
at the immediate neighboring point has completed.
However, as we describe in more detail below, the
integration of short characteristics along different rays may proceed
independently.  This means that one may expose parallelism by
choosing {\it angular decomposition} as an alternative.  In this
approach, a single processor has access to needed field variables
over the entire physical domain on only a subset of the angular mesh.
This avoids the problem with spatial recursion, but the
need to store (even a small number of) field variables for the
entire spatial domain on one processor places a practical limit
on total problem size.

On distributed-memory parallel architectures, data transfer between
processors is a necessity.  In codes which solve finite-differenced
forms of the basic equations, grid points along the local domain
boundaries on a given
processor will require information from the processors containing
grid points in the neighboring subdomains.  On distributed-memory
machines, data is shipped between processors in the
form of messages which are sent to designated target processors.
This type of communication is called {\it message passing}, and we
have adopted the
Message Passing Interface (MPI) standard to handle communication
in our code.

The decision to employ message-passing in a parallel code raises
perhaps the most critical requirement for efficient parallel
execution: overlapping computation with communication.  \zmp
~handles this requirement through the use of asynchronous message
passing, in which data send and receive operations proceed simultaneously
with program execution.  In this approach, the programmer must explicitly
ensure (through MPI WAIT or BARRIER constructs)
that data which is inbound to a local processor has actually arrived
before it is used.
This approach requires careful ordering of computation instructions
with respect to instructions for data communication; the basic
methods by which this done in~\zmp~are documented in \citet{fied97}.

In seeking to implement an efficient parallel scheme for the transport
solution, we have chosen to decompose the SC algorithms for $I$ along
the $\mu$ angular coordinate.
Recall that in the tangent-plane method, integration over the full
range of $\mu$ angles is performed in each tangent plane; more
importantly, the spatial integration for $I(\mu,\Phi)$ may proceed
independently for each value of $\mu$.  Decomposition of the $\mu$
coordinate is thus a natural means to achieve parallelism.  In
this method, each processor evaluates $I(\mu,\Phi)$ for a specified
subset of the discrete $\mu$ ordinates.

\begin{figure}
\plotone{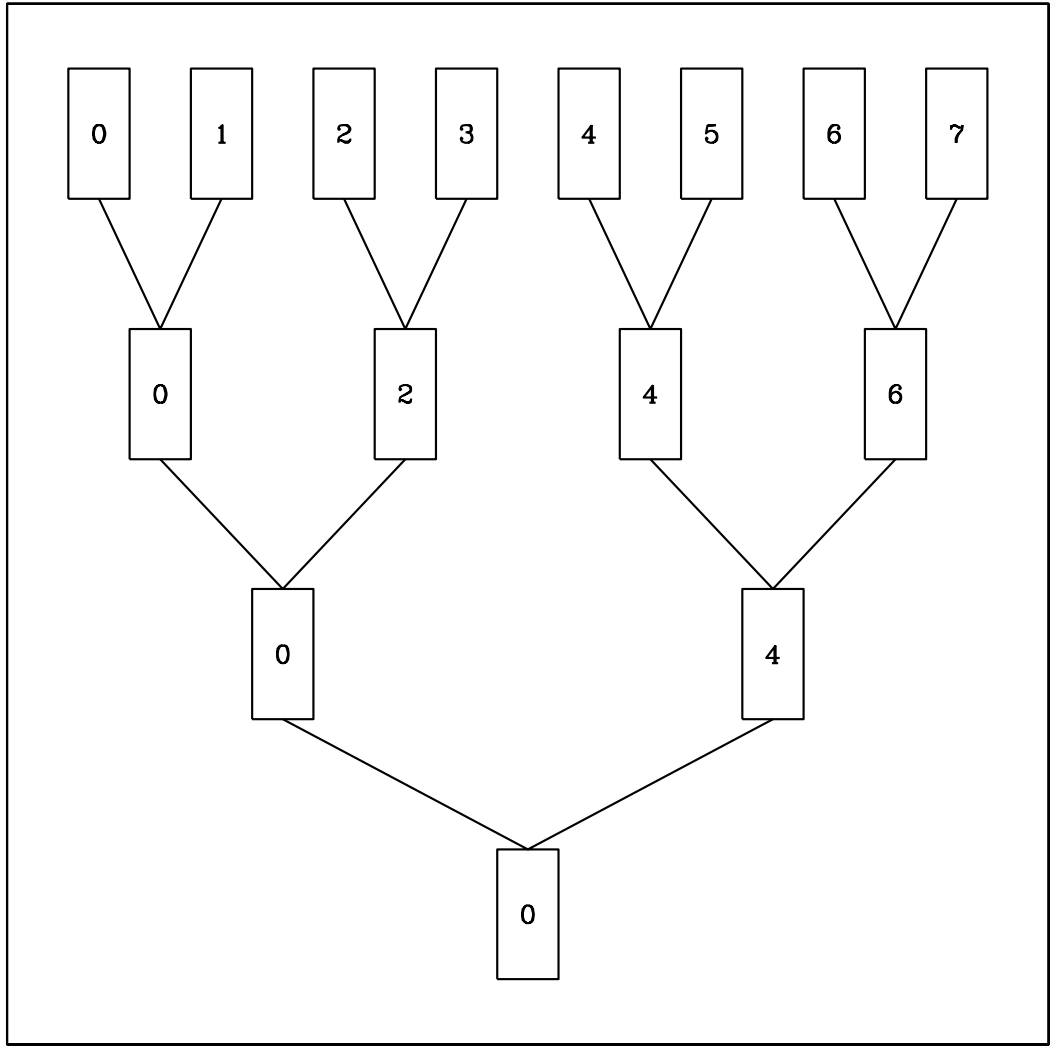}
\caption{An example of the data movement tree structure for a
global sum operation among eight processors.  Operations begin
at the top level, with neighboring pairs of rank (n-1,n)
combining data on processor (n-1).  The process proceeds down
each level until the root processor (0) has collected all the
data.  At a given level, the data movement within all processor
pairs proceeds in parallel.\label{fig2}}
\end{figure}

Once $I$ is known for each
subset, a global summation over all $\mu$ angles must be performed.
The result of this global sum must eventually be distributed to all
of the processors.
A global summation is by nature a serial operation, but it is possible
to partially parallelize the $\mu$ integral by dividing the
summation into a staged sequence of binary sums, illustrated by the
example in figure~\ref{fig2}.  Here, we see the method in the case of
an eight-processor parallel solution.  Initially, subsets of
$I(\mu,\Phi)$ are computed and stored on each processor.  Ultimately,
the grand sum is to be collected on the root processor (0) and
broadcast to the remaining processors.  In this example, the
integration proceeds in three communication stages, with processor
$n$ ($n > 0$) sending a local sum to processor $n-1$.  The
process cascades down levels until a complete sum is collected on
the root processor, which then broadcasts the result back to the
full process array.

Note that the procedure outlined here is repeated
on each tangent plane.  The integration over $\Phi$ occurs
incrementally as successive tangent planes are processed.  We have
not yet considered the question of whether an analogous degree of
parallelism can be performed over tangent planes in an efficient
manner.  If such a feature proves feasible, then time-dependent
transfer calculations on even larger grids than those considered
in this paper will be possible.  The construction of the $\Phi$
integration from tangent planes is sufficiently complex, however,
that we have elected to defer such research for future study.

\section{Numerical Test Problems}\label{tests}

The VTEF solution to the RHD equations requires a marriage of several
distinct physics modules and supporting numerical schemes.  Validation
of our approach will proceed in a systematic way, beginning with test
problems in static media.  In this category, we will first examine
tests of the radiation moment equations in which the Eddington tensor
is diagonal and characterized by a single scalar (either $1/3$ or 1)
which remains constant with time.  Both optically thick and thin
regimes will be considered.  We will then consider a two-dimensional
``shadow'' test wherein
the moment equations are solved with variable Eddington tensors.
Finally, we will present a series of radiation hydrodynamic tests,
utilizing both constant and variable Eddington tensors, exploring
the evolution of subcritical and supercritical radiating shocks.
In all of the numerical tests presented in this section, timestep
size is governed so that the maximum fractional change in the radiation
and gas energy densities ($\Delta E/E$ and $\Delta e/e$) from the
moment solution update is no more than five percent, and in the
case of the first problem that follows, this tolerance was lowered
to two percent.

 \subsection{Hydrostatic Moment Equations with Fixed Eddington Tensors}

   \subsubsection{Optically Thin Streaming: Plane Wave}

Our first problem tests the performance of the flux equation in
the free-streaming limit.  This one-dimensional test computes the
propagation of a plane wave through a medium of very low optical
depth, with no coupling to matter.   The purpose of this simple
test is to verify that the moment equations produce the correct
signal speed without recourse to a flux-limiter, and further to
gauge the precision to which a sharp wavefront may be resolved on
the computational grid.  We define a domain of length
1.0 cm, with a total optical depth of 0.01.  The radiation energy
density is initialized to $1.0\times 10^{-10}$ erg cm$^{-3}$ throughout
the domain; a value of 1.0 erg cm$^{-3}$ is specified at the inner
(left) boundary at t = 0.  Since this is a one-dimensional problem, we used a
large number (800) of zones along the Z axis to illustrate the 
best result our algorithm can achieve.  The Eddington tensor was
diagonal with constant values of (1,0) for ($\f_{11}$,$\f_{22}$).
We note also that, owing to very the very tight tolerance on allowed
changes in $\Delta E/E$ in a cycle, the timesteps remained extremely
low during the advance of the wavefront.  By the end of the run,
the timestep had only grown to a value of $2.3 \times 10^{-15}$ seconds,
an order of magnitude less than the radiation Courant time for the
problem.

\begin{figure}
\plotone{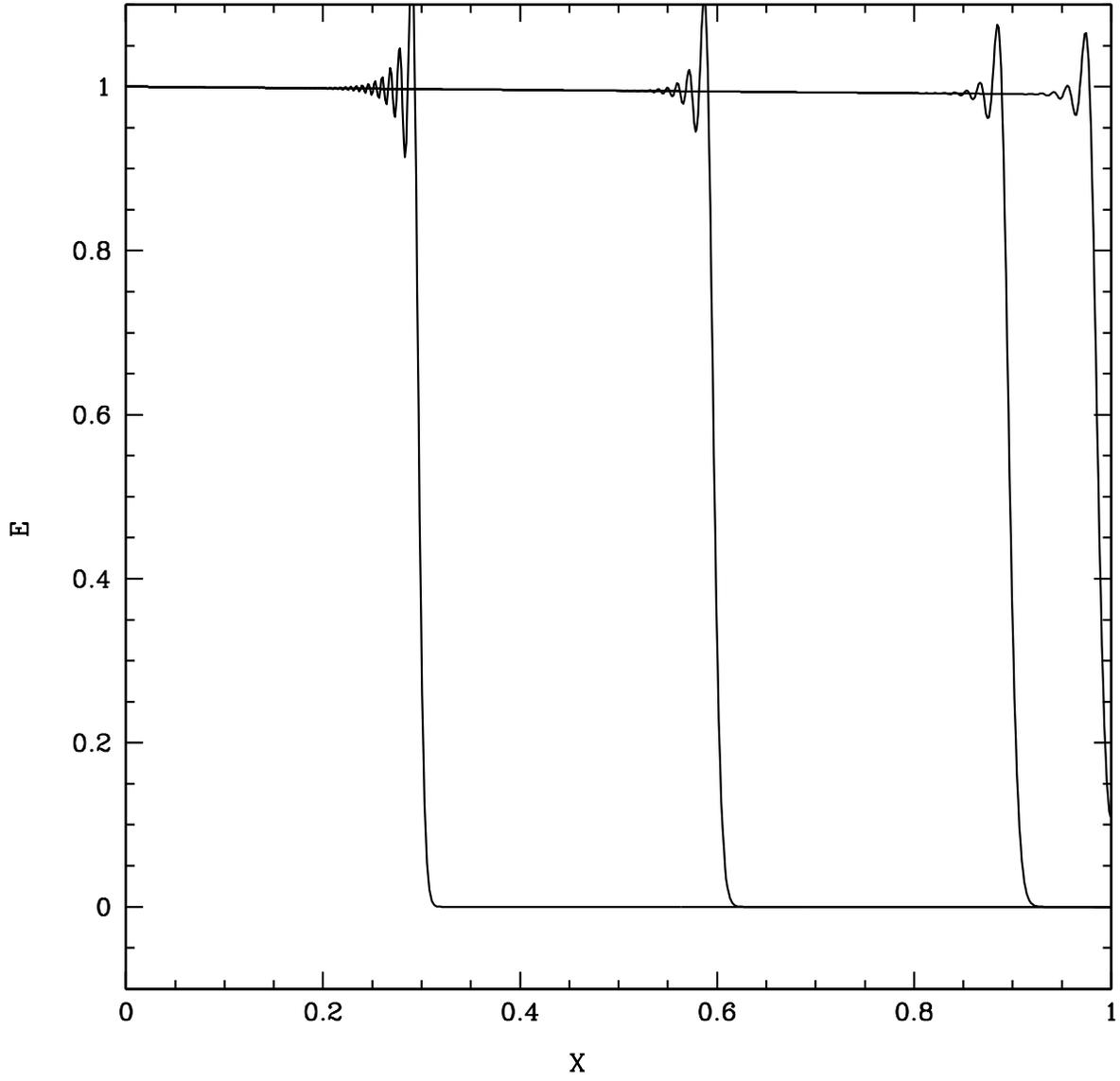}
\caption{Optically-thin streaming with a step-function source:
hyperbolic moment solution. \label{fig3}}
\end{figure}

\begin{figure}
\plotone{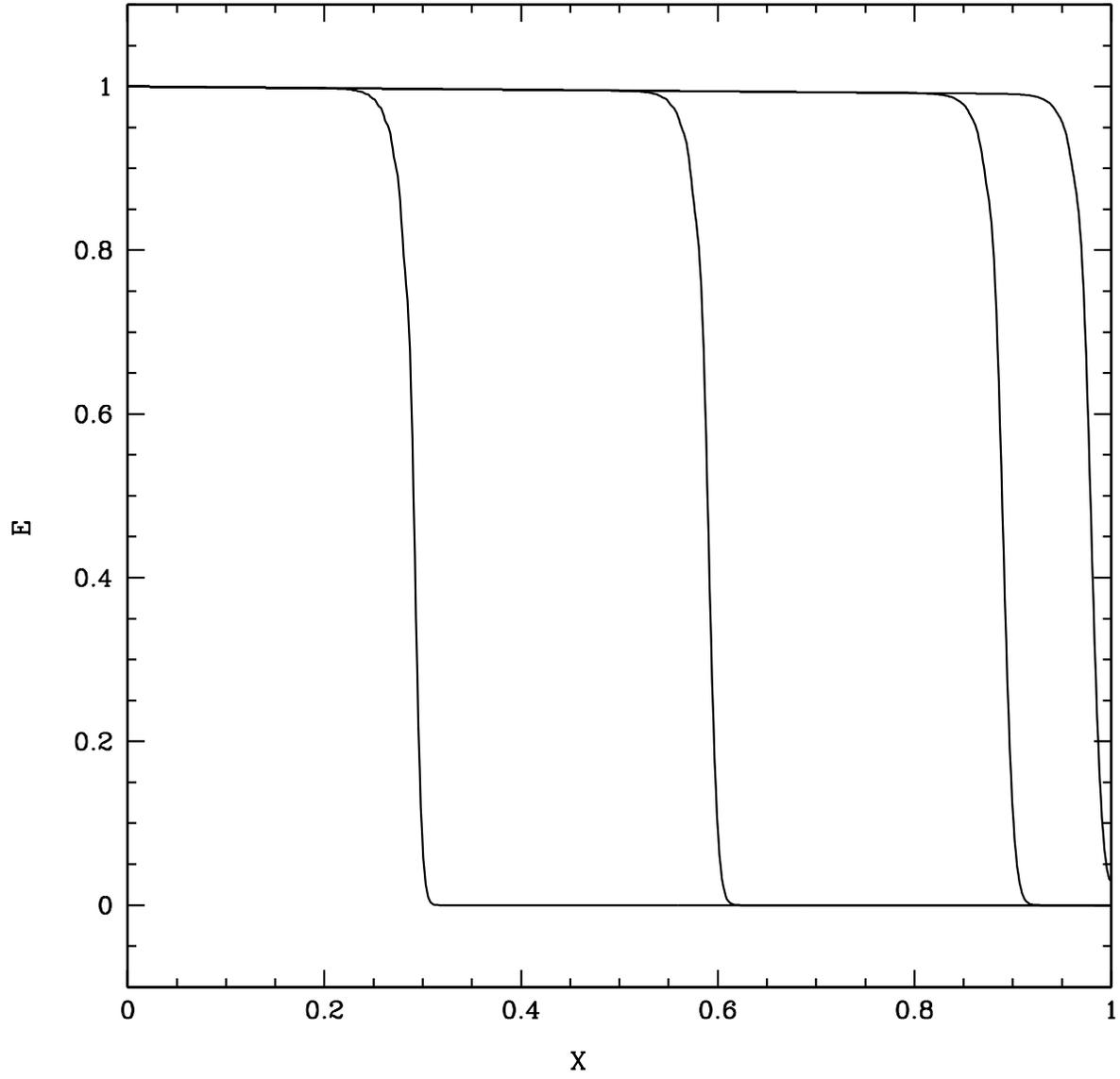}
\caption{Optically-thin streaming with a time-rounded source:
hyperbolic moment solution. \label{fig4}}
\end{figure}

Figures~\ref{fig3} and~\ref{fig4} show the positions of the wavefront
at $1.0\times 10^{-11}$, $2.0\times 10^{-11}$, $3.0\times 10^{-11}$,
and $3.3\times 10^{-11}$ seconds.  Figure~\ref{fig3} shows the result
for the case of an illuminating source turned on instantaneously at
t = 0.  This produces an infinitely steep wavefront; the oscillations
behind the front are a symptom of the hyperbolicity of our system
of moment equations.  
Figure~\ref{fig4} shows the result for the
same problem when the illuminating source is ramped up to its full
value in a very short time period.  In this case, we employ a
time-dependent boundary value, $E_{bv}(t)$, for the radiation energy
density:
\begin{equation}
E_{bv}(t) \ = \ E_{\infty}\left(1 - e^{(-t/t_{o})}\right).
\label{radbc}
\end{equation}
In our simulation, $t_{o}$ was taken to be 9 times the light-crossing
time for an individual zone.  $E_{\infty}$ is again equal to 1.0.
We see that the wavefront is still
very well resolved, and that it moves across the grid with the correct
signal speed.  We are encouraged to see that a very steep wavefront
can be stably propagated without recourse to a more elaborate numerical
scheme; while it may be possible to propagate a truly discontinous
radiation wave front, the astrophysical need for such an ability is
dubious at best.

      \subsubsection{Diffusion and Matter Coupling: Marshak Waves}

We add matter-radiation coupling to our test suite through
the solution of a non-equilibrium diffusion Marshak wave problem.
The problem formulation is identical to that described in \citet{su96},
who fashioned their test after that described by \citet{pom79}.
This idealized case is characterized by a
purely absorbing, semi-infinite medium initially at zero temperature.
In order to design a problem with an analytic solution, the matter
is characterized by a single opacity which is independent of 
temperature; furthermore, the matter has a specific heat
proportional to the temperature cubed.  This results in a set of
equations which are linear in $E$ and $T^{4}$.  Pomraning defined
dimensionless space and time coordinates as
\begin{equation}\label{xvar}
x \ \equiv \ \sqrt{3}\kappa z,
\end{equation}
and
\begin{equation}\label{tvar}
\tau \ \equiv \ \left({4ac\kappa \over \alpha}\right) t,
\end{equation}

and introduced dimensionless dependendent variables, defined as
\begin{equation}\label{uvar}
u(x,\tau) \ \equiv \ \left(c \over 4\right) \left[E(z,t) \over
F_{\rm inc}\right],
\end{equation}
and
\begin{equation}\label{vvar}
v(x,\tau) \ \equiv \ \left(c \over 4\right) \left[aT^{4}(z,t) \over
F_{\rm inc}\right].
\end{equation}

In (\ref{uvar}) and (\ref{vvar}), $F_{\rm inc}$ is the incident
boundary flux.  With the definitions given by (\ref{xvar})
through (\ref{vvar}), Pomraning showed that the radiation and gas
energy equations could be rewritten, respectively, as
\begin{equation}\label{ueq}
\epsilon{\partial u(x,\tau) \over \partial \tau} -
{\partial^{2} u(x,\tau) \over \partial x^{2}} \ = \ 
v(x,\tau) - u(x,\tau),
\end{equation}
and
\begin{equation}\label{veq}
{\partial v(x,\tau) \over \partial \tau} \ = \ 
u(x,\tau) - v(x,\tau),
\end{equation}
subject to the following boundary conditions:
\begin{equation}
u(0,\tau) - {2 \over \sqrt{3}}{\partial u(0,\tau)\over \partial x}
\ = \ 1,
\end{equation}
and
\begin{equation}
u(\infty,\tau) \ = \ u(x,0) \ = \ v(x,0) \ = \ 0.
\end{equation}

In (\ref{ueq}), $\epsilon$ is related to radiation constant and specific
heat through
\begin{equation}
\epsilon \ = \ {4a \over \alpha}.
\end{equation}

\begin{figure}
\plotone{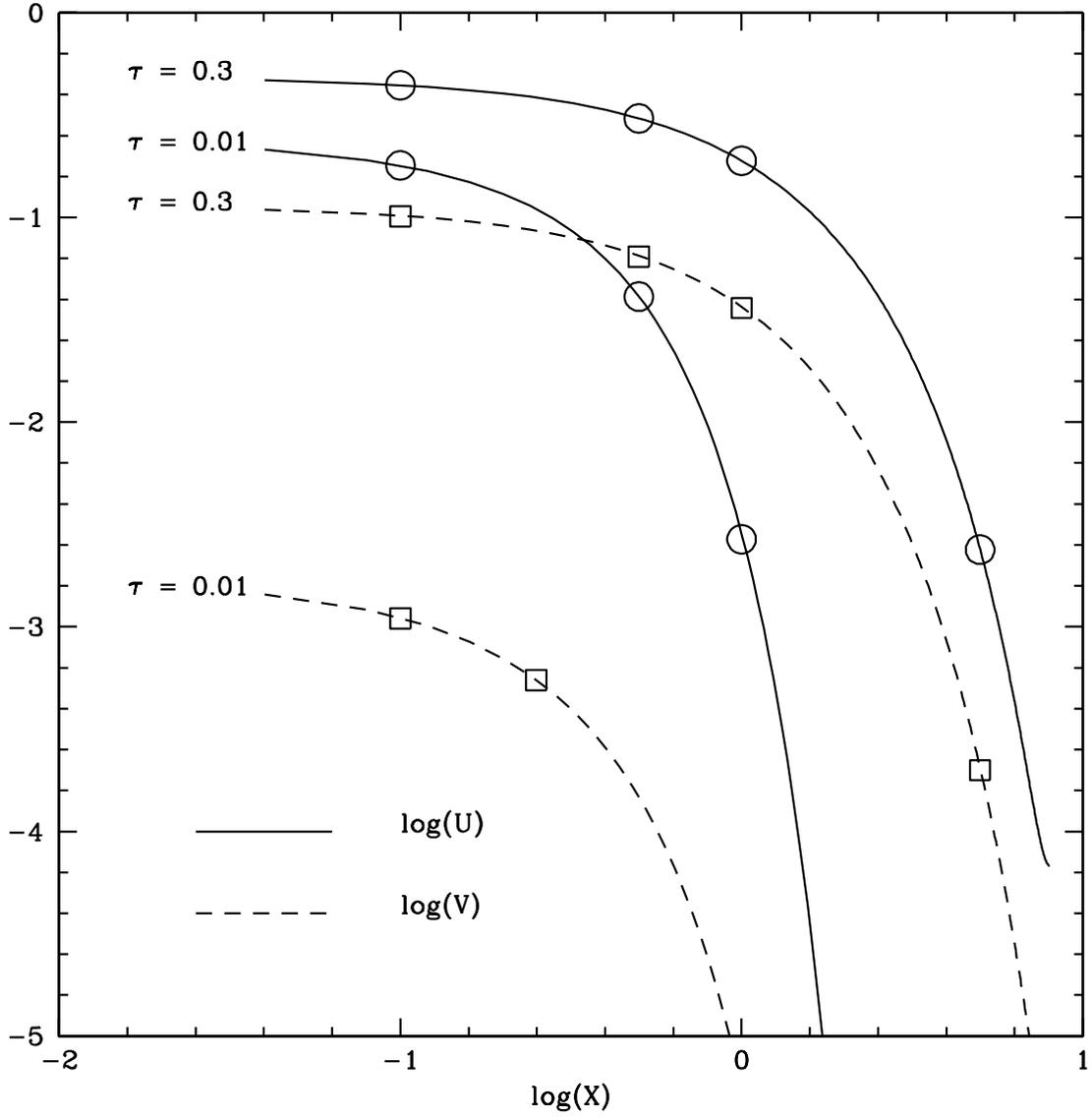}
\caption{Non-linear Marshak wave: Numerical solutions (curves)
and analytic solutions (circles/squares) \label{fig5}}
\end{figure}

Once epsilon is specified, the problem is characterized and amenable
to both analytic and numerical solution.  In their 1996 paper, Su
and Olson published a set of benchmark results for a range of epsilon
parameters.  In addition to the simple forms of the opacity and
specific heat, this problem assumes pure diffusion; i.e. there is
no flux limiting employed.  In order for our moment solver to be
deployed on this problem in a meaningful way, the time-derivative
terms in the radiation flux equation (which guarantee flux limiting
automatically) must be artificially zeroed.  With such restrictions,
this problem has little utility from
a transport perspective, but it does provide a useful check on the
operator splitting scheme we use to perform matter-radiation coupling.
Our Eddington tensor is diagonal
with each element equal to $1/3$.
Figure~\ref{fig5} shows the results of a test performed for an
$\epsilon$ value of 0.1.  The curves represent numerical data;
the circles and squares are taken from analytic solutions published
by \citet{su96}.  The agreement is excellent.

   \subsection{Hydrostatic Moment Equations with VTEF: Casting
               Shadows}

This test examines what is perhaps the defining feature
of the VTEF method: the ability to reproduce and preserve strong
angular variations in the radiation field.  While we will see
that there are some limitations of the success of our method, the
qualitative difference between the VTEF results and those from 
FLD is so striking, with the VTEF results much closer in line
with physical expectations, that we regard these results as a
major step forward.  Our problem consists of a spheroidal, opaque
cloud which is irradiated on one side by a distant source.  In
the transfer equation solver, the specific intensity is subject
to a beam-like boundary condition of the following form:
\begin{equation}
I_{\rm bndry} \ = \ I_{\rm src}\delta(\mu) \ + \ I_{\rm amb}.
\label{beambndry}
\end{equation}
This results in forward-peaked radiation
consistent with a point-source object at a great distance from the
target.  The cylindrical domain is 1.0 cm in length and 0.12 cm in
radius.  The ambient density within the cylinder is $10^{-3}$ gm
cm$^{-3}$; an oblate spheroid of density 1.0 gm cm$^{-3}$ is aligned
along the symmetry axis, with its center located at (Z,R) =
(0.5,0).  

The shape and density structure of the cloud were determined by the
following:
\begin{equation}
\rho_{\rm cloud}(z,r) \ = \ \rho_{0} \ + \ {\left(\rho_{1}-\rho_{0}
 \right) \over {1 + \exp{\Delta}}},
\label{cloudsurf}
\end{equation}
where
\begin{equation}
\Delta \ = \ 10\left( \left({z(i,j)-z_{c}} \over z_{0}\right)^{2} + 
                      \left(r(i,j)       \over r_{0}\right)^{2}
 \ - \ 1 \right).
\label{cloudshape}
\end{equation}
In (\ref{cloudshape}), $z_{c}$ is the axial coordinate of the
cloud center, and $(z_{0}, r_{0})$ equal (0.10,0.06).  Equation
(\ref{cloudshape}) both defines a rotational ellipsoid and 
in conjunction with equation
(\ref{cloudsurf}) imparts a ``fuzzy'' surface to the cloud in the
sense that the density does not transition from the exterior
value ($\rho_{0}$ = 0.001 gm cm$^{-3}$) to the interior value
($\rho_{1}$ = 1.0 gm cm$^{-3}$) instantaneously.

In the first test presented in this section, we examined the case
where a nearly discontinuous profile in the radiation energy density
was propagated, assuming a constant isotropic Eddington tensor.
In that instance,
the radiating source was fully illuminated almost instantaneously.
Experience acquired during the development of this
code showed that such a choice is problematic when truly two-dimensional
problems are considered.  The difficulty with quasi-sharp wavefronts
is associated with limitations of the transport solution to track
very rapid time variability in the radiation, a point which we 
discuss in detail in \S{\ref{discuss}}.
For now, we note that the radiation energy equation was subjected
to a time-dependent boundary condition of the form shown in equation
(\ref{radbc}), with a characteristic time, $t_{o}$, equal to one
light-crossing time for the full length of the cylinder.  This results
in an advancing radiation wave with a very broad wavefront.  The
characteristic temperature of the source at full illumination was
0.15 eV, or approximately 1740 K.  The opacity was chosen such that
\begin{equation}
\chi \ = \ \chi_{0}\left(T \over T_{0}\right)^{-3.5}\left(\rho \over
\rho_{0}\right)^{2},
\end{equation}
with $T_{0}$ equal to the initial uniform domain temperature, and
$\rho_{0}$ with the value given above.  $\chi_{0}$ was chosen to
be 0.1 cm$^{-1}$, ensuring a nearly transparent medium outside the
target and a highly opaque medium below the cloud ``surface.''

\begin{figure}
\plotone{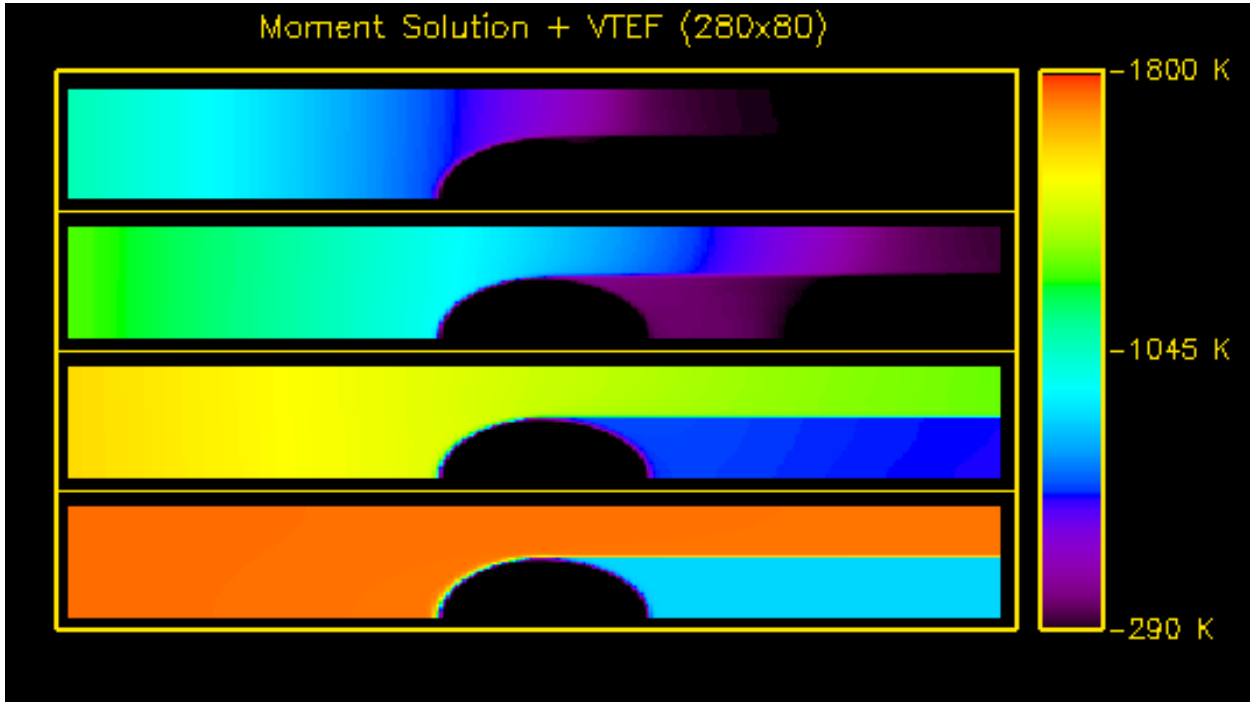}
\caption{Shadow test with VTEF:
the passage of a plane wave over an opaque spheriod.  From top
to bottom, the panels correspond to 0.68, 1, 2, and 10 light-crossing
times.  The moment variables and Eddington tensors were both computed
on a 280x80 (ZxR) grid.
\label{fig6}}
\end{figure}

\begin{figure}
\plotone{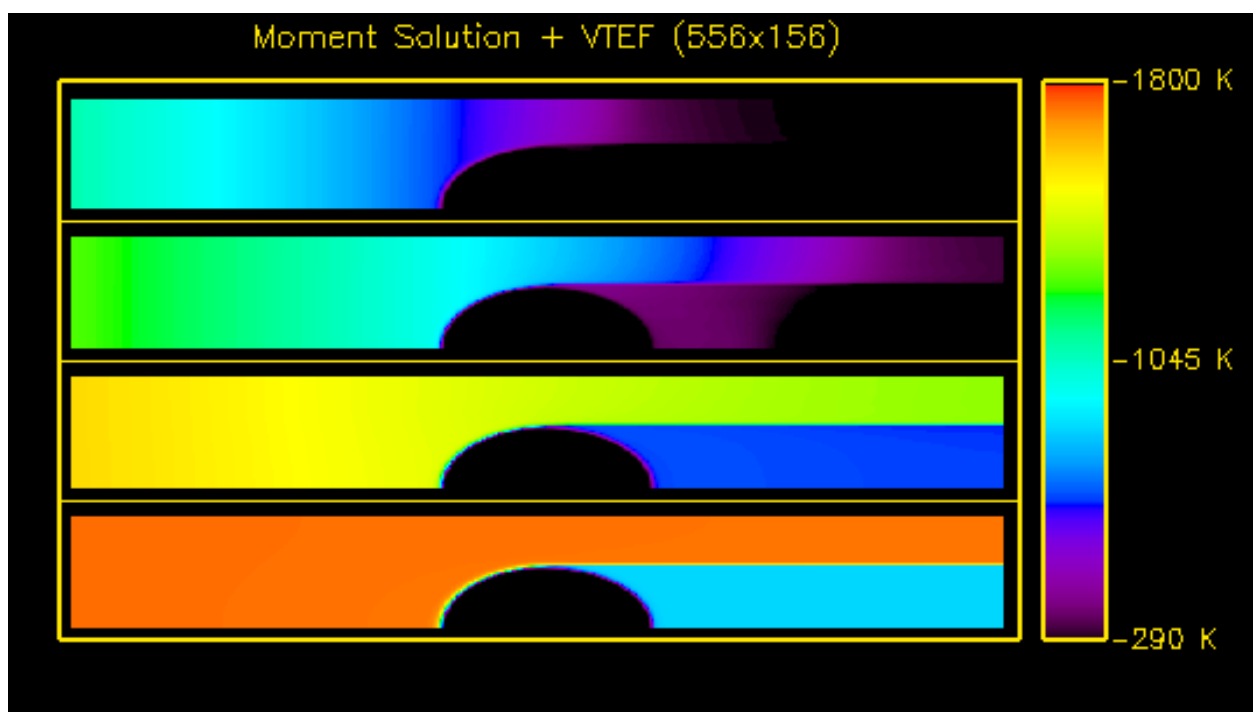}
\caption{Shadow test with VTEF:  as in figure~\ref{fig6}, but
for a 556x156 moment grid combined with a 280x80 transfer solution
grid.
\label{fig7}}
\end{figure}

\begin{figure}
\plotone{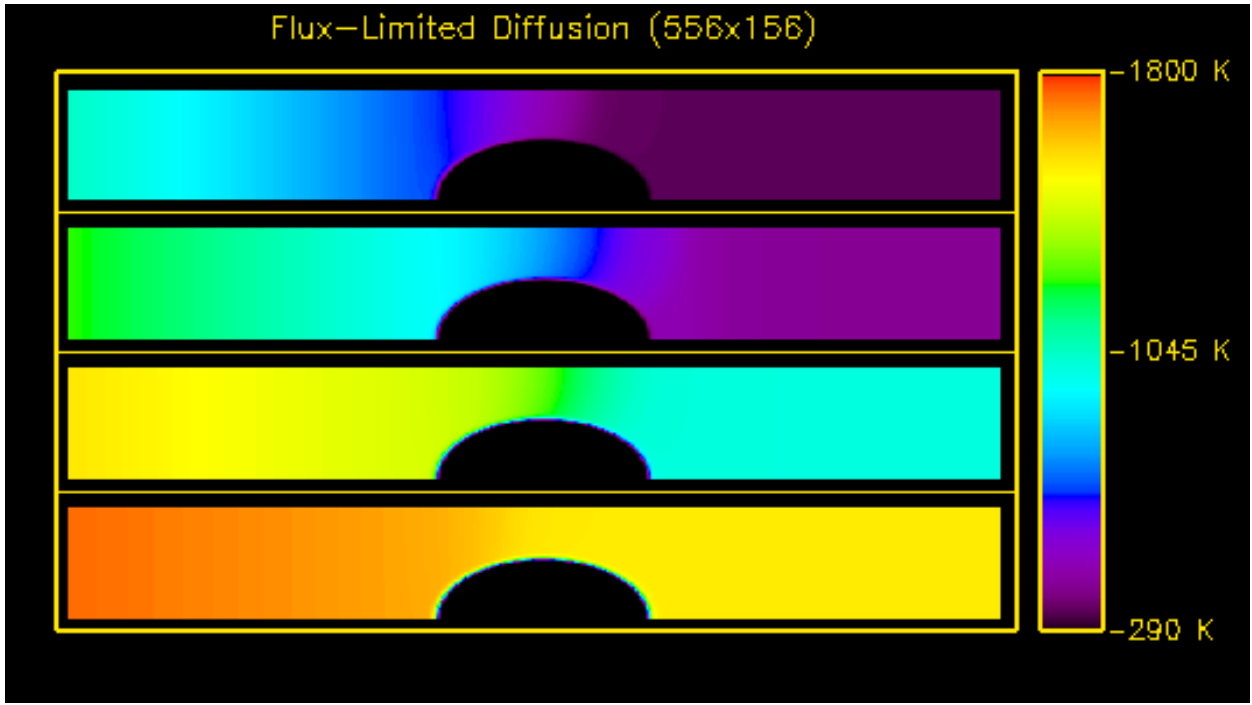}
\caption{Shadow test with FLD: a flux-limited diffusion comparison
test at 556x156 resolution.  The times are the same as for the
VTEF cases.  Note that the cloud shadow is completely destroyed.
\label{fig8}}
\end{figure}

\begin{figure}
\plotone{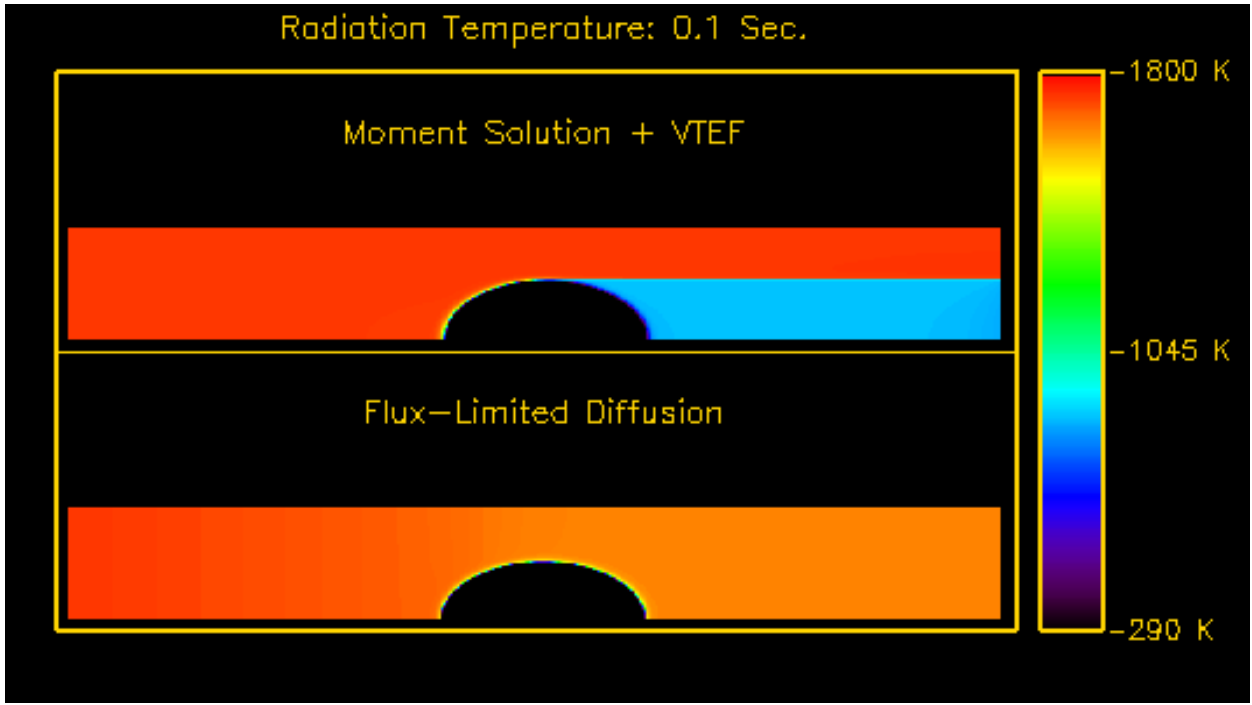}
\caption{VTEF vs. FLD: the radiation energy density at 0.1 seconds
($3 \times 10^{9}$ light-crossing times) for the VTEF calculation
(top) and the FLD calculation (bottom).  Note that the VTEF calculation
has remained essentially unchanged from its asymptotic state in
figure~\ref{fig7}.\label{fig9}}
\end{figure}

As defined, this problem was evolved for 0.1 seconds, which is
$3 \times 10^{9}$ light-crossing times for the cylinder.  Figures
\ref{fig6} through \ref{fig8} present images of the radiation
temperature distribution throughout the domain during the first
ten light-crossing times of the simulation.  Figure~\ref{fig6}
gives results for a simulation in which the radiation moment variables
and the transfer solution were evolved on a 280x80 (ZxR) grid.
The PSTAT algorithm was used to compute Eddington tensors; 33
$\mu$ angle cosines were distributed over 16 CPU's, with three
angles stored on the root process and two angles on each of the
remaining processes.  Four panels showing the full ZxR domain are
shown, corresponding to 0.68, 1.0, 2.0, and 10 light-crossing times.
To begin with, we see the advance of the broad radiation wave, and
the definition of a shadow region as the wave passes over the
spheroidal target.  As the source becomes fully illuminated, the
shadow remains clearly delineated, but there is some leakage of
radiation from the illuminated region into the shadow.  Because
we assume no scattering in the problem, and further because neither
the limb of the target nor the low-density gas have yet warmed
enough to re-emit significantly, this leakage is not physical.  It
is due, rather, to the diffusivity of the radiation energy equation
which depends upon a ``double divergence'' of the Eddington tensor
(cf. equations (\ref{rade2}), (\ref{radm0}), and Appendix A).
Once the radiation field reaches the state shown in the fourth
panel, however, it remains stable with energy leaked into the shadow
being transported out the exit boundary.  The radiation field at
three billion light-crossing times is virtually unchanged from its
state at ten light-crossing times (figure~\ref{fig9}).

Figure~\ref{fig7} presents the same calculation, assuming the
same physics and employing the same transport techniques, for a
radiation moment grid at 556x156 resolution.  In this run, the
moment grid is divided into a 4x4 process topology, and a sampling
parameter ($N_{\rm samp}$) of 2 was used to define a CSC grid on
which Eddington tensors were computed.  Physically, the results
are virtually indistinguishable, which shows that the diffusivity
in the moment equations is not affected by higher resolution.  The
root of the problem is not resolution; it is instead a consequence
of the fact that fluxes in the radial direction are spuriously
generated by a finite-difference stencil that ``straddles''
boundaries between illuminated and shadowed regions.

While we have identified a significant weakness in the current
VTEF implementation, we have not yet moved to remedy it.  Instead,
we proceed to examine results from the traditional approach to
transport: flux-limited diffusion.  Figure~\ref{fig8} presents
a high-resolution (556x156) run in which the radiation field is
evolved according to FLD:
\begin{equation}
\F \ = \ -{c\over\chi}\Lambda(E)\Delta E.
\label{fld}
\end{equation}
In (\ref{fld}), $\Lambda(E)$ is the flux-limiter; we have used the
Levermore-Pomraning limiter for the results given here.

\begin{figure}
\plotone{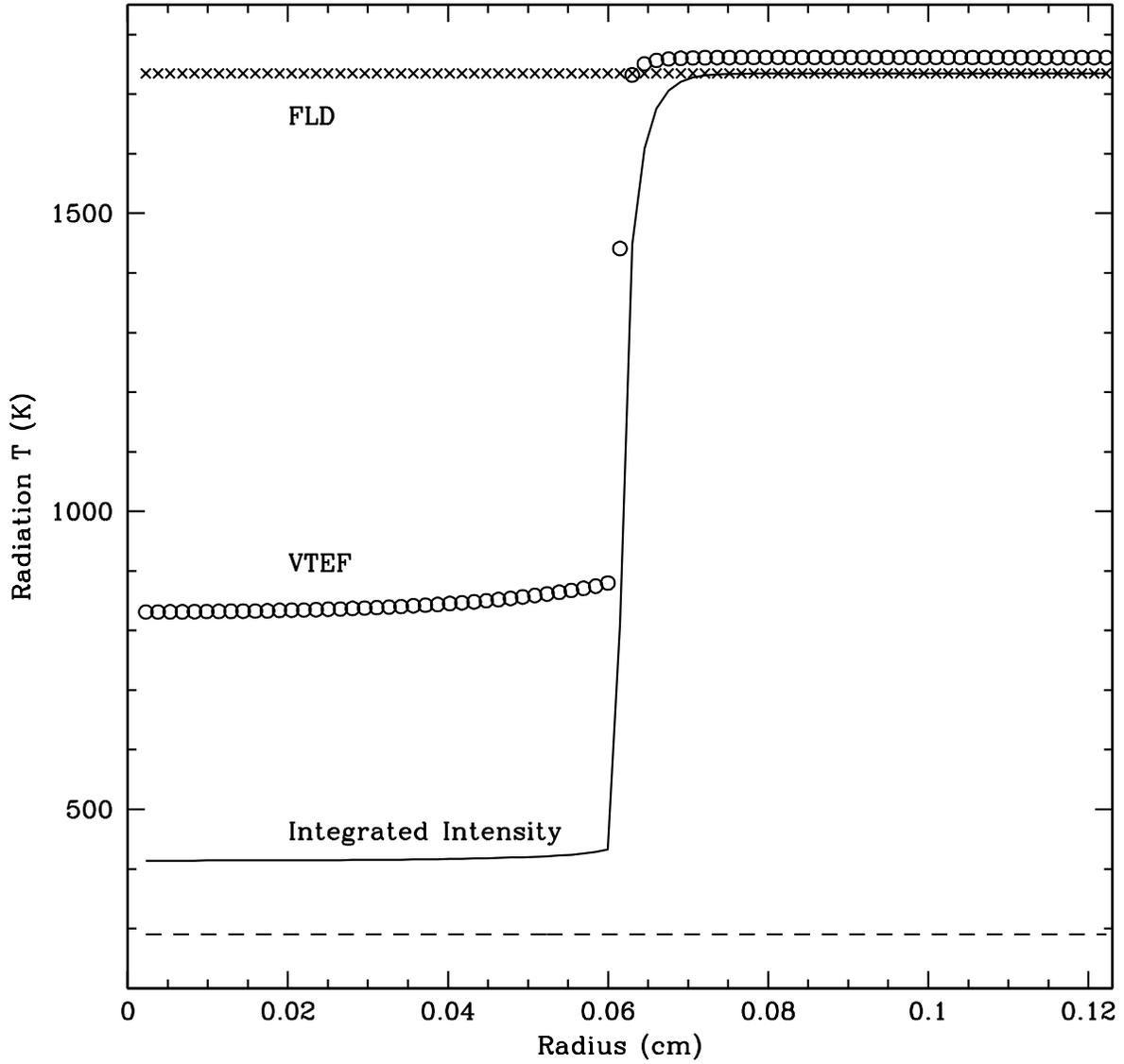}
\caption{Radial profiles of radiation temperature: computed values
of T$_{rad}$ from FLD (crosses) and VTEF (open circles), compared
to that derived by integrating the specific intensity over angle
(solid line).  The temperature at t = 0 is indicated by the dashed
line.\label{fig10}}
\end{figure}

Even after only 10 light-crossing times, the difference between
FLD and VTEF is dramatic and fundamental.  In the FLD prescription,
radiation flows around and surrounds the target as rapidly as it
moves down the cylinder.  No delineation of a shadow is ever
present, and at long times (figure~\ref{fig9}) the target is
irradiated isotropically, even though the illuminating source is
defined to exist at only one end of the cylinder!  Thus while we
acknowledge room for improvement in the numerical implementation
of VTEF, we see already a qualitative difference in the two approaches
which will have fundamental consequences when full hydrodynamic
simulations are considered.  We conclude this section with an
examination of figure~\ref{fig10}, which shows the variation of radiation
temperature with radius at $Z$ = 1.0 cm (the center of the obstructing
cloud is at $Z$ = 0.5 cm) at the end of the evolution (0.1 sec).
The dashed line indicates the initial
profile prior to the passage of the radiation front.  Open circles
show the radiation temperature computed from a full VTEF calculation;
open crosses show results from an FLD calculation.  The solid line
shows radiation temperatures derived from the zeroth moment of the
specific intensity extracted from our transport solver.  This is
the same quantity used to derive the Eddington tensors for the moment
solution, but notice that the radiation energy derived from $I$
does not exhibit the large, spurious leakage of energy into the shadow
region.  $I$ has risen, within the shadow, above its initial ambient
value slightly because of thermal re-emission from low-density gas in
the non-shadowed region of the domain.  This also contributes (slightly)
to the filling in of the shadow in the VTEF case, but there is a large
component of this which is artificial.  Nonetheless, the VTEF algorithm
has managed to preserve a well-defined shadow even after $3 \times
10^{9}$ light-crossing times, a task at which FLD fails completely.

   \subsection{Radiation Hydrodynamics: Radiating Shocks}

In our final collection of test results, we marry the radiation
transport algorithms to the gas hydrodynamic modules of \zmp.
Our test problem in this instance is the evolution of radiating
shock waves in optically thick media.  The presence of radiation
modifies the shock structure considerably, introducing a radiative
precursor which preheats the material downstream from the shock
to some characteristic temperature, $T_{-}$.  Denoting the temperature
behind the shock front as $T_{2}$, we may follow the discussion in
\citet{mih84} and identify two major classes of radiating shocks:
{\it subcritical} and {\it supercritical} shocks.  In the subcritical
case, \ttwo~is greater than \tm, and the radiation precursor is
relatively weak.  At higher flow velocities, \tm increases relative
to \ttwo, and at some critical velocity $u_{\rm crit}$, the two become
equal (\tm never exceeds \ttwo).  Shocks for which $u \geq u_{\rm crit}$
are termed supercritical.  As shall be seen below, supercritical
radiating shocks are characterized by a very large radiation precursor
which heats the unshocked material downstream from the shock front.

      \subsubsection{Subcritical Shocks}

Our problem configuration is guided by that published by \citet{ens94},
who considered a battery of test problems designed for astrophysical
codes.  Ensman considered the case of a piston moving through static
media, and followed the evolution with a one dimensional, Lagrangean,
radiation hydrodynamics code (VISPHOT).  Because \zmp~evolves the gas dynamic
equations in the Eulerian frame, we pose the problem analogous to
the Noh test where a moving medium impacts a stationary reflecting
boundary.  Velocities in our tests are initialized to match the
piston speeds specified in the Ensman paper.  Since VISPHOT assumes
a spherically symmetric problem, Ensman generated a (nearly) planar
problem by placing the medium in a thin shell at large radii.  Since
our grid is cylindrical, we can produce a rigorously planar shock,
and we choose problem dimensions, initial densities, and opacities
consistent with those published in \citet{ens94}, save for one
feature: the Ensman grid attempts to reproduce an infinite plane.
Our cylindrical grid has a finite radius which is smaller than the
axial length of the domain, which means that our test involves a
radiating surface which is formally finite in extent.  This effects
the degree of forward peaking in the radiation field downstream from
the shock, and accounts for some qualitative differences between our
profiles and those published by Ensman.

In our subcritical test, we define a cylinder of length $7.0\times
10^{10}$ cm, with an initial uniform density of $7.78\times 10^{-10}$
g ${\rm cm}^{-3}$.  The gas and radiation temperatures (initially
in equilibrium) were set to 10K at the outer boundary and increased
progressively by 0.25K in each interior zone.  This choice was made
by Ensman to avoid numerical difficulties with zero flux in VISPHOT;
we repeat the practice for consistency.  This problem assumes a
purely absorbing medium with a constant opacity; our value was
$3.1 \times 10^{-10}$ cm.  The physical domain is therefore almost
22 times greater than the photon mean-free path.  Thus, while the
medium may be characterized as optically thick {\it in toto}, individual
zones are optically thin, a point we will consider further in the
supercritical case.

The problem is initiated by setting the axial velocity equal to
$-6.0 \times 10^{5}$ cm s$^{-1}$ everywhere on the grid.  300 uniform
zones were used to span the Z axis, whereas a modest 16 zones spanned
the radial direction, itself only $3.2 \times 10^{9}$ cm in extent.
The domain, then, assumes the shape of a long, narrow pipe, with
variations in physical quantities confined to the Z direction.

In order to produce figures which are directly comparable to results
obtained with lagrangean codes such as VISPHOT, the ``Z'' axis shown
in the following plots has been transformed to show the positions
in a frame in which an unshocked parcel of matter still moving
with the inflow velocity, $v_{i}$, is at rest.  Thus, if $Z_{i}$ denotes
the plot coordinate, and $Z_{o}$ is the lab frame coordinate, we have
\begin{equation}
Z_{i} \ = \ Z_{o} - v_{i}t.
\end{equation}
In this way we may represent our results as if they had been generated
by a piston in a stationary medium.

\begin{figure}
\plotone{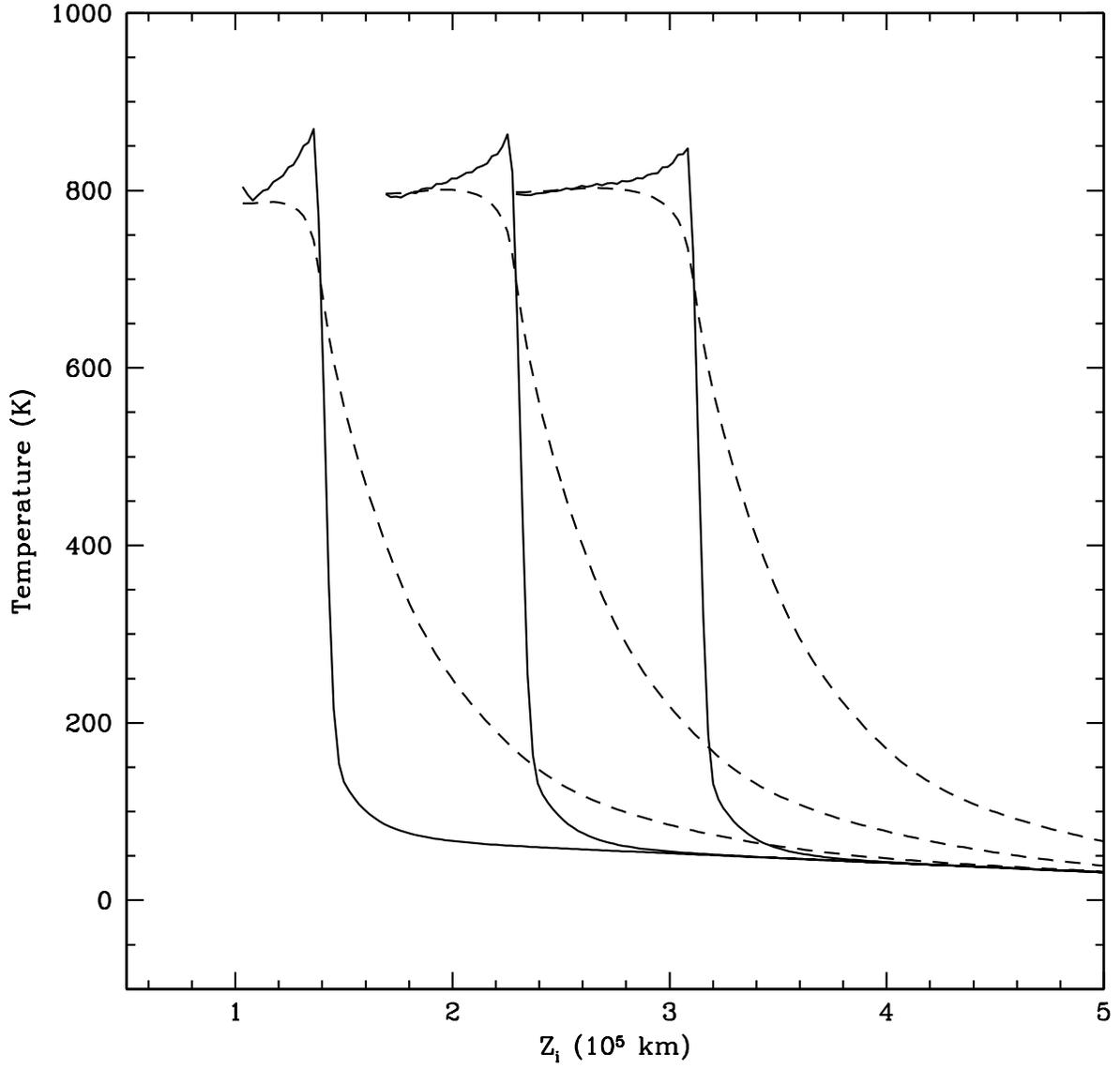}
\caption{Subcritical Shocks: gas (solid line) and radiation (dashed
 line) temperatures at $1.7\times 10^{4}$,
 $2.8\times 10^{4}$, and $3.8\times 10^{4}$ seconds. \label{fig11}}
\end{figure}

\begin{figure}
\plotone{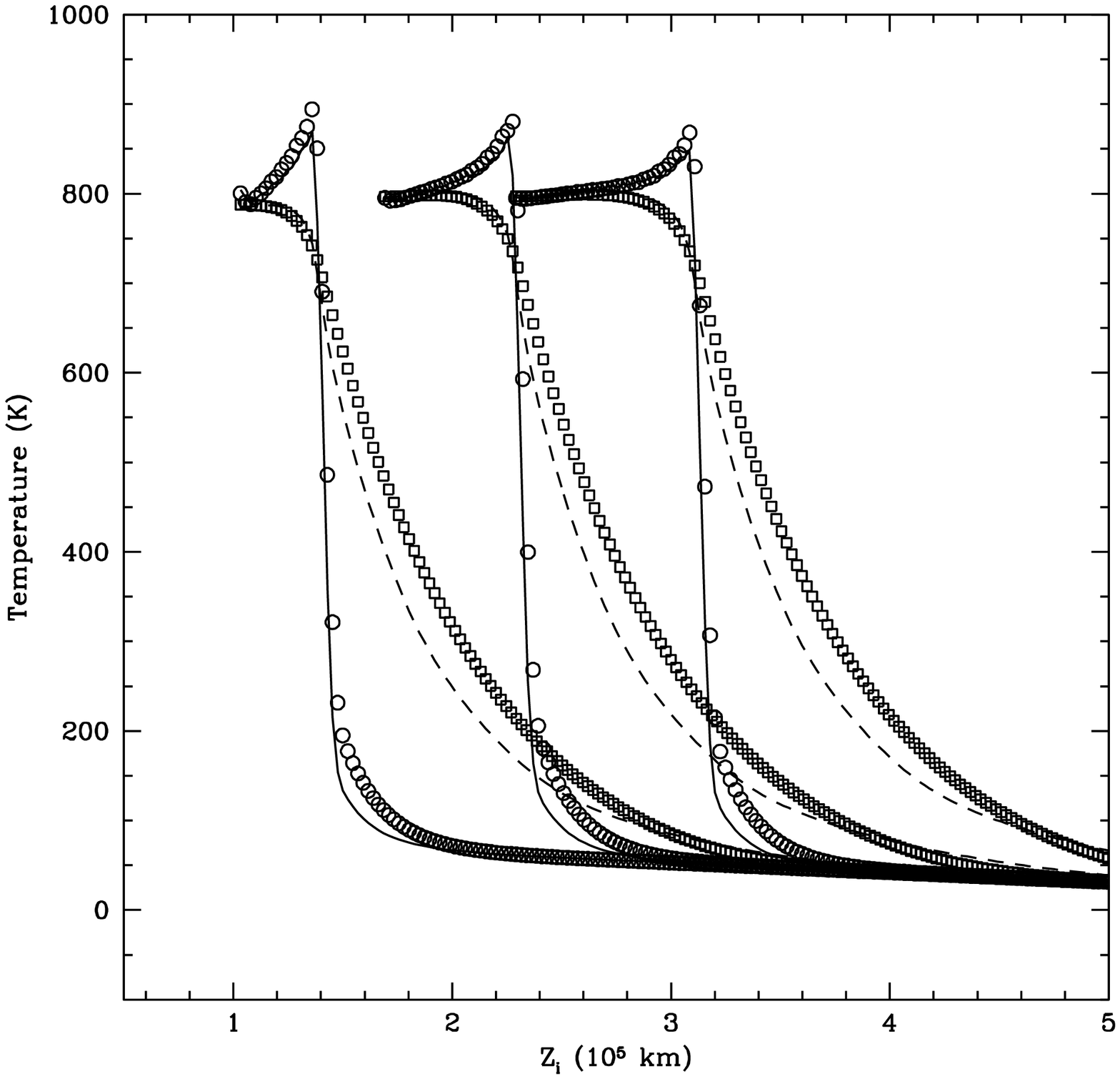}
\caption{Subcritical Shocks: Temperature profiles computed with
VTEF-PSTAT algorithm (curves), compared to those computed assuming
a constant, isotropic Eddington tensor (circles and squares).
The times are the same as those in figure~\ref{fig11}.
\label{fig12}}
\end{figure}

Figure~\ref{fig11} shows profiles for the radiation temperature
(solid lines) and gas temperature (dashed lines) at 3 fiducial
times in the evolution.  In this test, the VTEF transport algorithm
was not employed.  Instead, a constant, isotropic Eddington tensor
with values of 1/3 along the diagonal was chosen to close the radiation
moment equations.  We refer to this choice as the ``CTEF'' (constant
tensor Eddington factor) approximation.
In figure~\ref{fig11}, the radiation precursor
is clearly evident, the matter and radiation temperatures are well
out of equilibrium on both sides of the shock front.
Figure~\ref{fig12} presents results for the same problem configuration
with a full VTEF closure employed in the radiation moment equations.
In this case, the PSTAT time-dependent solution to the transfer
equation was chosen to calculate the Eddington tensor.  The closure
values were updated whenever the gas temperature at some point in
the domain changed by at least two percent.  In this test, we see
that the difference between the VTEF and CTEF results is
relatively minor.  The radiation profiles are slightly broader in
the CTEF case, but the positions of the advancing shock front are
identical in both cases, as is the matter temperature variation
across the shock.  This result is perhaps not surprising given that
radiation effects are weak in the subcritical case, and the physical
domain is optically thick.  Furthermore, we note that our solutions
are consistent with the subcritical solutions shown in figure 8 of
Ensman (1994).

      \subsubsection{Supercritical Shocks}

In the supercritical shock tests, all physical parameters remain
the same as in the subcritical case except for the inflow velocity,
which has been changed from -6 to -20 km/sec along the Z-axis.
As before, the problem is initialized on a 300x16 (ZxR) grid.
In plotting profiles of physical quantities at various times, we
repeat the practice of transforming Z values into the inertial
frame in which the inflowing material is at rest.  

\begin{figure}
\plotone{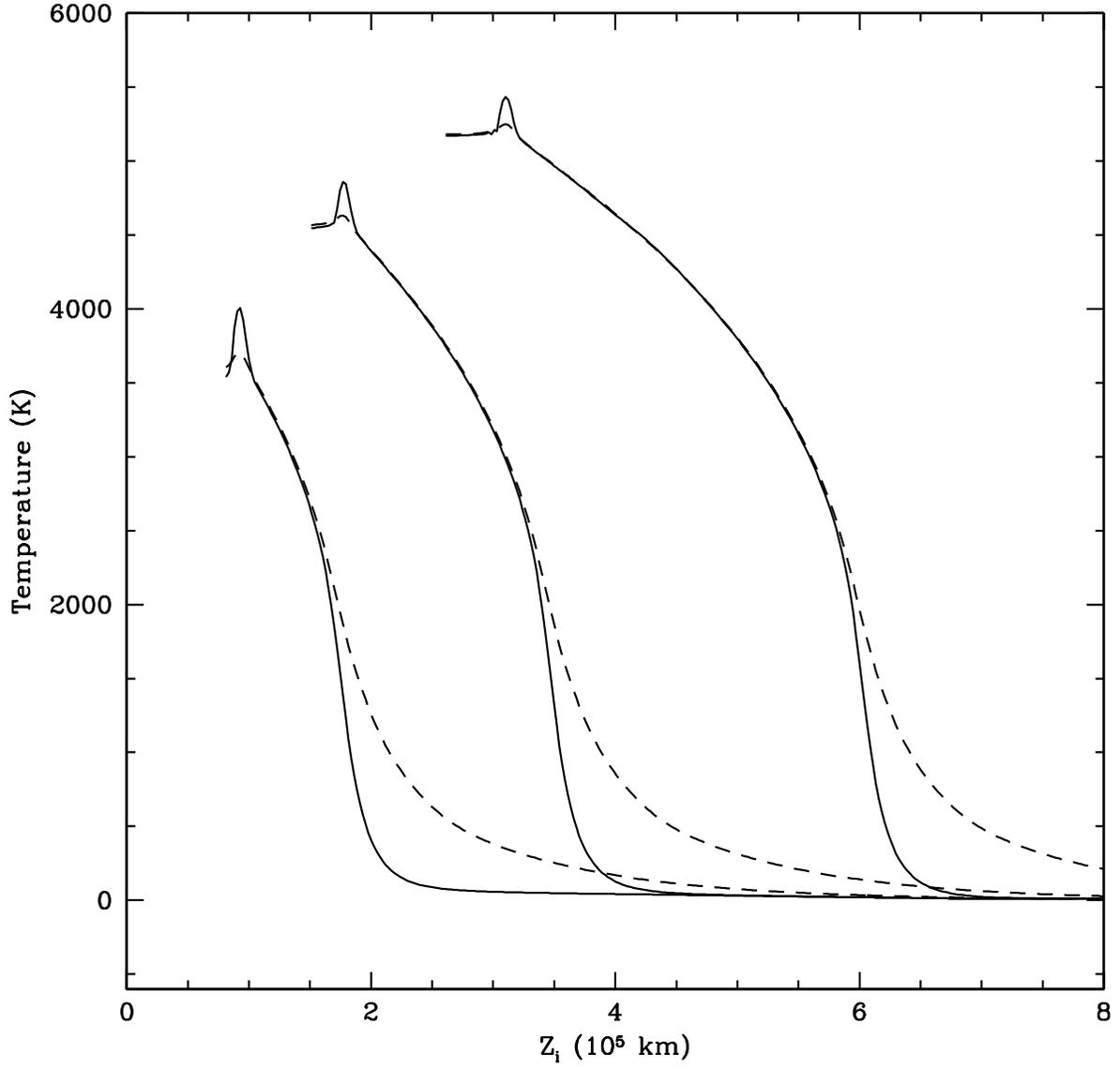}
\caption{Supercritical Shocks: gas (solid line) and radiation
 (dashed line) temperatures at $4.0\times 10^{3}$,
 $7.5\times 10^{3}$, and $1.3\times 10^{4}$ seconds. \label{fig13}}
\end{figure}

In contrast with
the subcritical test, we present as our standard result a model
in which the full VTEF-PSTAT algorithm is employed.  Figure
\ref{fig13} shows profiles in gas and radiation temperature
(solid and dashed lines, respectively) for three times in the shock
evolution.  The radiative precursor has become a dominant feature
of the preshock medium.

\begin{figure}
\plotone{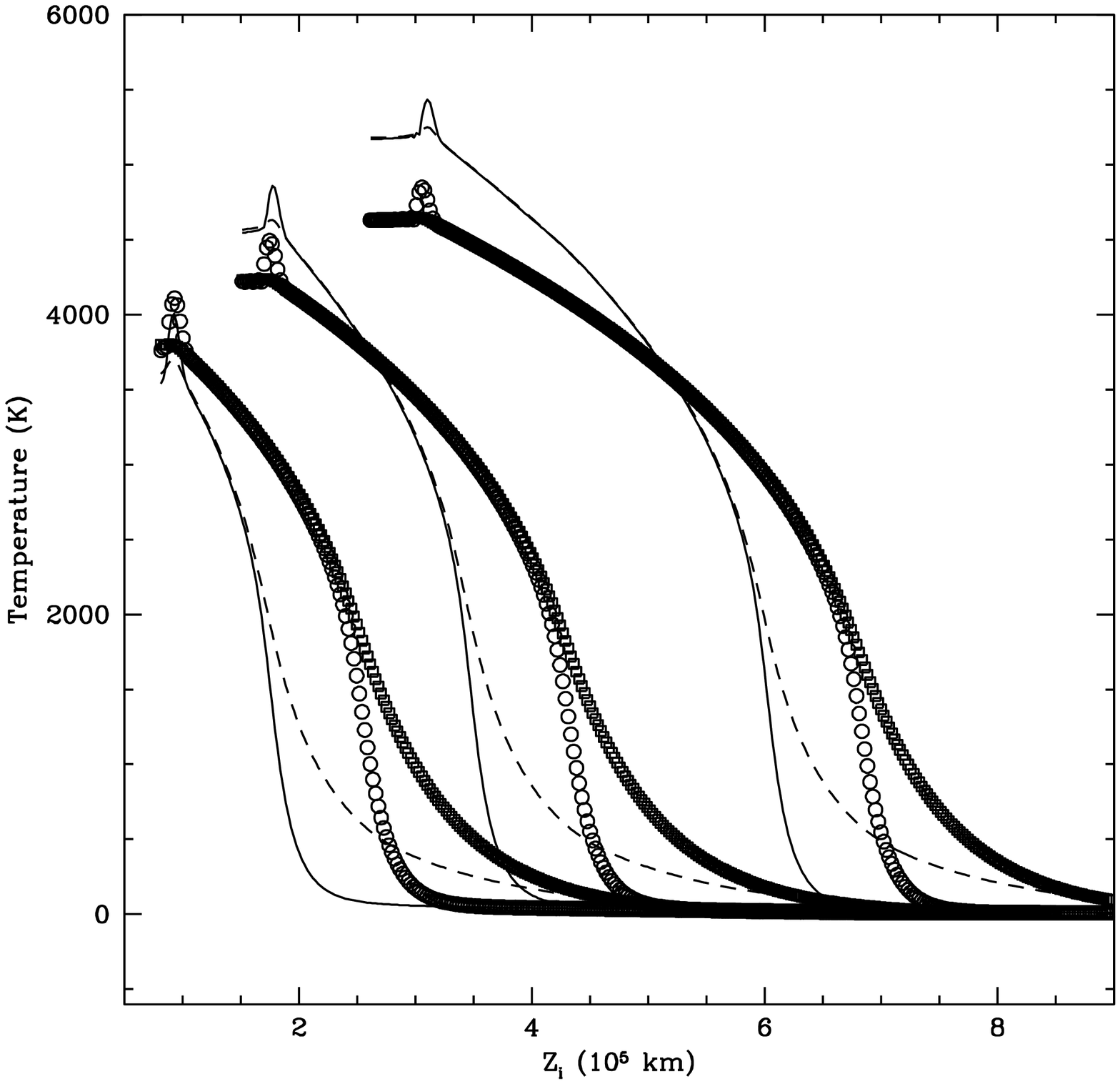}
\caption{Supercritical Shocks: Temperature profiles computed with
the VTEF-PSTAT algorithm (curves),
compared to those computed assuming
a constant, isotropic Eddington tensor (circles and squares).
The times are the same as those in figure~\ref{fig13}.
\label{fig14}}
\end{figure}

\begin{figure}
\plotone{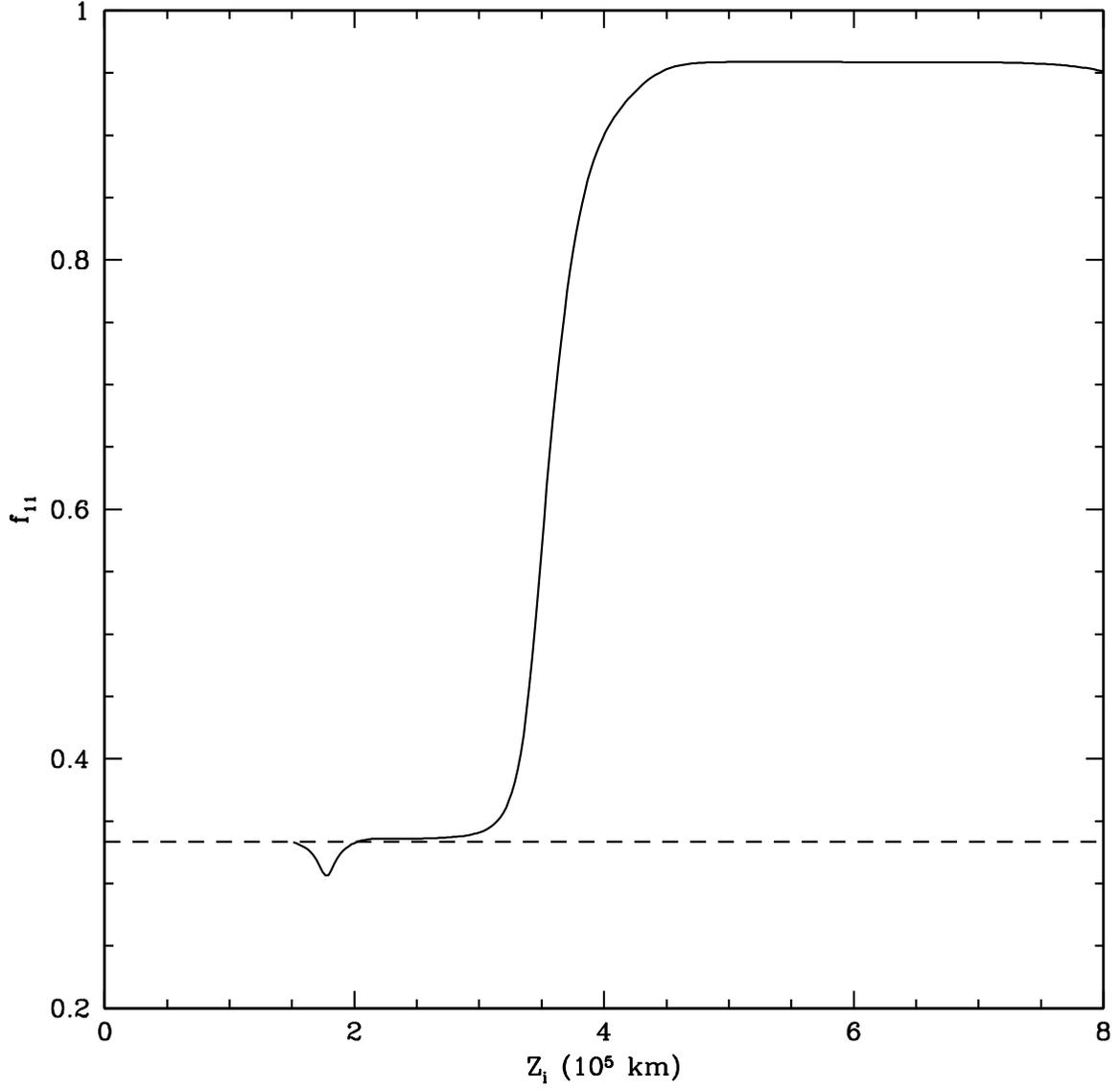}
\caption{Supercritical Shocks: The \f$_{11}$ Eddington
tensor component, at t = $7.5\times 10^{3}$ seconds,
for the run shown in figure~\ref{fig13}.  The dashed line is located
at a value of 1/3.  Note the dip below 1/3 at the
position of the shock front, a known signature of radiating
shocks. \label{fig15}}
\end{figure}

The significant role played by radiation in the supercritical case
suggests that the details of transport physics will be
more important than in the subcritical case.  We affirm this
expectation by presenting a comparison calculation performed with
a constant Eddington tensor containing values of 1/3 along the
diagonal (our CTEF approximation). Figure~\ref{fig14}
combines the gas and radiation temperature profiles from
figure~\ref{fig13} with data from the CTEF run.  The difference
is fairly dramatic, with the CTEF run yielding profiles which are
broader but lower in amplitude.  The position of the shock front
with time is unchanged, as is the relative height and width of
the non-equilibrium temperature spike at the shock front.
Figure~\ref{fig15} show, for the time at which the middle profile
is plotted, the spatial run of the \f$_{11}$ Eddington tensor
component.  In the vicinity of the shock, high temperatures on
both side ensure a relatively isotropic radiation field, but well
downstream \f$_{11}$ approaches the streaming limit of 1, even
though the medium is (globally) optically thick.  Of further
interest is the slight dip below 1/3 at the position of the shock
front.  This is a real effect which is characteristic of a strong
transverse component in the specific intensity at that point.

That the results vary strongly between the CTEF and VTEF solutions
is a major point of departure between our caculations and the
corresponding solutions in figure 15 of Ensman (1994).  We remind the
reader, however, of a significant geometric difference between the
two cases:  the Ensman model, being defined on a thin spherical
shell at a large radius, approximates a thin plane of infinite
extent in the directions perpendicular to the propagation vector.
Our calculation, however, is initialized on a long, narrow pipe.
The radiating surface we compute is therefore finite and results
in a radiation field which becomes very forward-peaked (cf.
fig.~\ref{fig15}).  The analytic value for the scalar Eddington
factor located at the surface of an infinite plane is 1/2, which
is much closer to an isotropic value than that seen in our calculations.
It is therefore not surprising that our supercritical shock tests
exhibit a stronger sensitivity to the transport physics than those
performed by Ensman.  In the subcritical case, in which radiative
effects are minor, the details of transport are considerably
less important.

\begin{figure}
\plotone{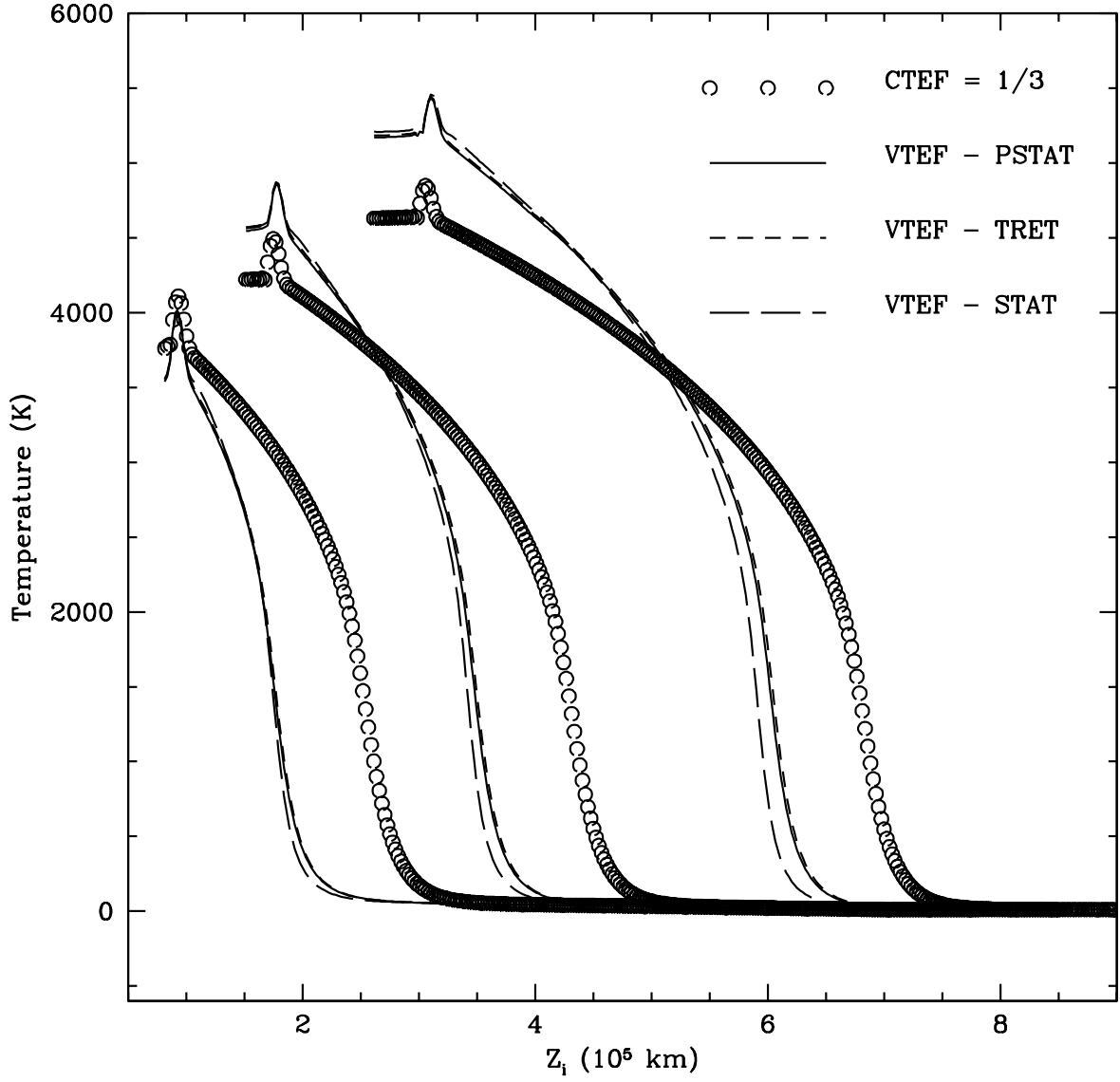}
\caption{Supercritical Shocks: A comparison of gas temperature
profiles resulting from all four Eddington factor calculations.
\label{fig16}}
\end{figure}

Figure~\ref{fig16} shows that, for this problem, the specific choice by which variable
Eddington tensors are computed is less important than the decision
to allow them to vary at all.  Because this problem is optically
thick, the radiation field changes on time scales which are long
compared to the light-crossing time (roughly 2 seconds).  Thus there
is very little gained by using either time-dependent form of the 
VTEF solution (TRET or PSTAT) in place of a purely static solution.
We note, however, that this statement would not hold were one to 
model a medium marked by a transition from optically thick to thin
media, such as a stellar surface (in the case of photon transport)
or the iron core of a collapsing supernova progenitor (in the case
of neutrino transport).

\section{Summary and Discussion}\label{discuss}

In this paper, we have presented a new code for performing RHD
simulations on parallel computers.  The algorithms discussed here
include and augment the functionality of the serial algorithms 
documented in~\citet{sto92c}. We have documented both the analytic
and discrete forms of the radiation moment solution and the
Eddington tensor closure term.  We have described three different
methods for computing a short-characteristic formal solution to the 
transfer equation, from
which our VTEF closure term is derived.  Two of these techniques
include time dependence, a feature not typically associated with
the formal solution in astrophysics literature.  Two key numerical
components of our implementation have been highlighted: the biconjugate
gradient linear system solver, written for general use on massively
parallel computers, and our techniques for parallelizing both the
radiation moment solution and the transfer solution.  Finally, we
have presented a suite of test problems which run the gamut from
optically thin transport with a fixed Eddington tensor, to full 
RHD+VTEF calculations of radiating shocks.

This document and our code possess a number of features which are
new with respect to Paper III and the serial code described therein.
In the moment equations, we chose to accompany the radiation energy
equation with an equation for the gas energy rather than adopting a
total energy equation.  This choice allows us to employ a different
matter-radiation coupling scheme with three very attractive features:
(1) it is particularly robust in regimes where matter and radiation
are far out of equilibrium, (2) it allows the construction of a moment
solution matrix involving only one dependent field variable, and (3)
it is extremely well-suited to implementation on parallel platforms.
In the transfer solution, we have retained the original static
method, suitable for computing Eddington tensors for static or 
quasistatic radiation fields, but we have added two algorithms which
include time dependence in different ways.  One of these treats
temporal effects by time-retarding the opacities and source functions
encountered along a characteristic ray.  The second of these, which
we have adopted as our default method for problems needing a 
time-dependent treatment, builds a discrete form of the
time derivative operator directly into the characteristic solution.
This ``pseudostatic'' form has an advantage over the ``time-retarded''
form in that it is not necessary to save opacities and emissivities
from a previous timestep.  All three algorithms may be used to compute
Eddington tensors on the full moment solution grid, or upon a subset
of mesh points obtained by subsampling the moment grid at regular
intervals along both coordinates.

Design for parallel execution is an entirely new feature with respect
to the older code.  We have used the Message-Passing Interface (MPI)
standard exclusively in constructing our code for parallel use.
In the gas dynamical and radiation moment solution modules, we
have distributed work among parallel processors via spatial domain
decomposition.  In contrast, the transfer solution has been subdivided
over the $\mu$ angle coordinate, owing to the spatially recursive
nature of the SC solution.

This work also contains a more diverse set of test problems than
that provided in Paper III.  We feel that this is particularly
important, because the RHD modules in Paper III have remained
unused since they were written more than ten years ago, due in part
to the subsequent research pursuits of their creator, and also to the
fact that the publicly available version of~\ztwd~contained an FLD
module rather than the VTEF algorithm described in Paper III.
As a result, the viability of this VTEF approach
for multidimensional calculations has not been properly documented
in the astrophysical literature.  As a first step toward correcting
this shortfall, we have provided full RHD tests to which the serial
algorithm was never subjected.  In addition, we have focused
considerable attention on multidimensional radiation fields containing
a shadowed region, consistent both the original mission of this
LLNL-funded project and with anticipated astrophysics applications
to come. 

The test problems we have examined allow us to consider the strengths
and weaknesses of the VTEF method for multi-dimensional problems.
The major weaknesses of the method, as currently implemented, are the
diffusivity of the discrete moment equations and limitations upon the
extent to which the transport solution can track rapid time variation
of the radiation field.  The diffusivity of the moment solution is
a direct consequence of the difference stencil used to discretize
the moment equations; the ``double divergence'' term in the radiation
energy equation (see equations \ref{rade3} and \ref{g1term}) results
in finite differences which artificially connect regions in space
on either side of physical boundaries in the radiation field, such
as the shadow edges in our illuminated cloud problem.  In our test
case, this prevents the VTEF algorithm from reproducing such shadowed
regions as fully as we would like.  Nonetheless, a noticeable shadow
is maintained for very long timescales, with leakage of energy
(via the R-component of the flux) into the shadowed region balanced by
transmission (via the Z-component of the flux) of energy through
the exit boundary. 

The second problem affects the degree of consistency between the
moment solution and transport solution for the radiation field.
Mathematically, our radiation moment equation results from a zeroth
angular moment of the time-dependent equation of transfer; using
a VTEF closure for the moment equation is therefore accurate only
if the radiation solutions as characterized by these two equations
are mathematically consistent.  An equivalent statement is that the
radiation energy density produced from the moment equation should
equal that obtained through a direct integration of the specific
intensity over all solid angles.  The first test problem in
\S\ref{tests} shows that our moment solution is capable of following
a sharply defined radiation front when the Eddington tensor is an
idealized constant with $\f_{11}$ = 1.  Our formal solution, however,
includes time dependence in an approximate manner, and disagrees
strongly with the moment solution when the radiation field shows
a rapid variation in time (and therefore space).

\begin{figure}
\plotone{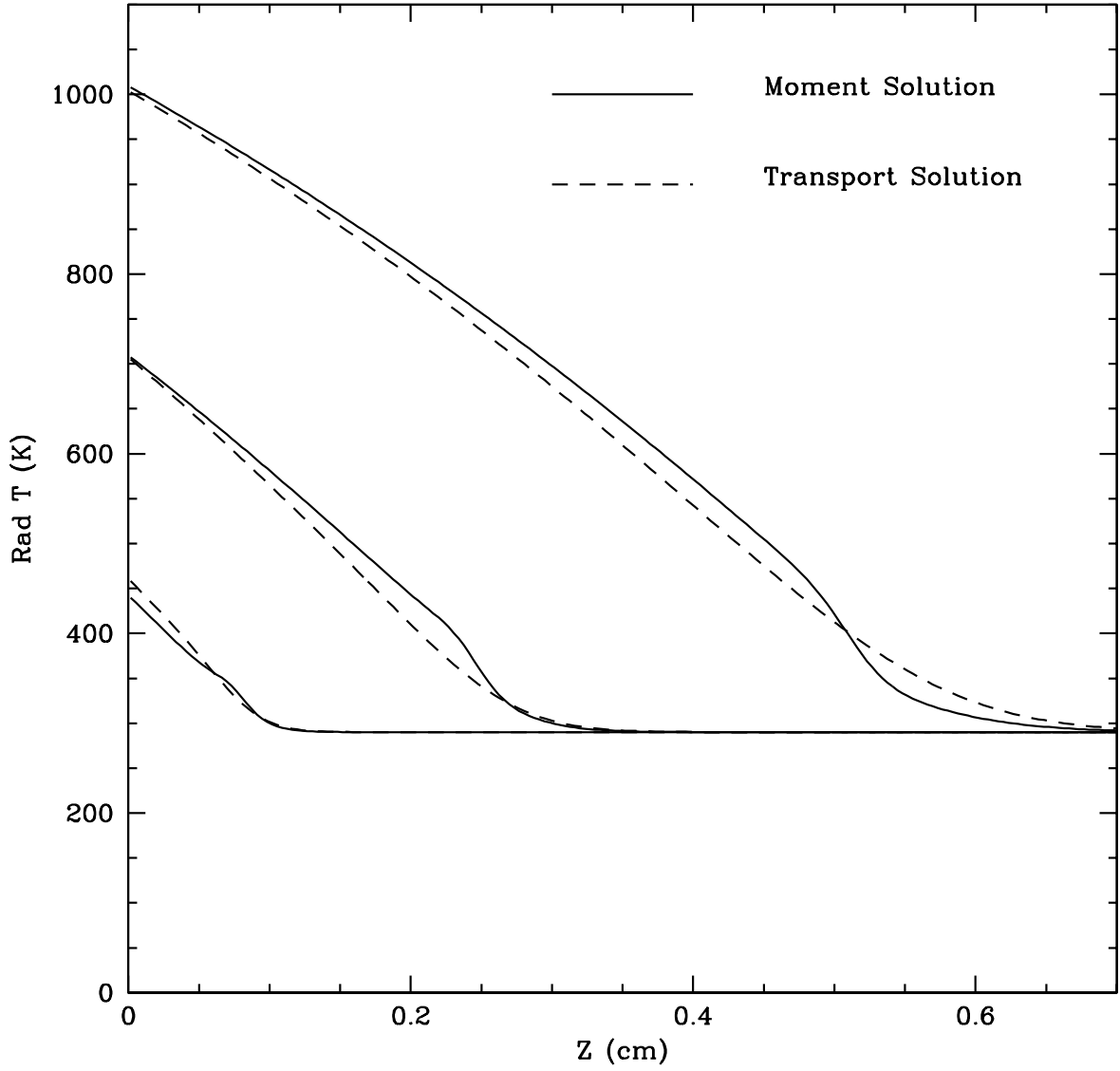}
\caption{Radiation temperatures from the moment solution (solid
line) compared to those from the transfer solution (dashed line)
for the case of a slowly varying illuminating source incident on
a uniform medium.
\label{fig17}}
\end{figure}

\begin{figure}
\plotone{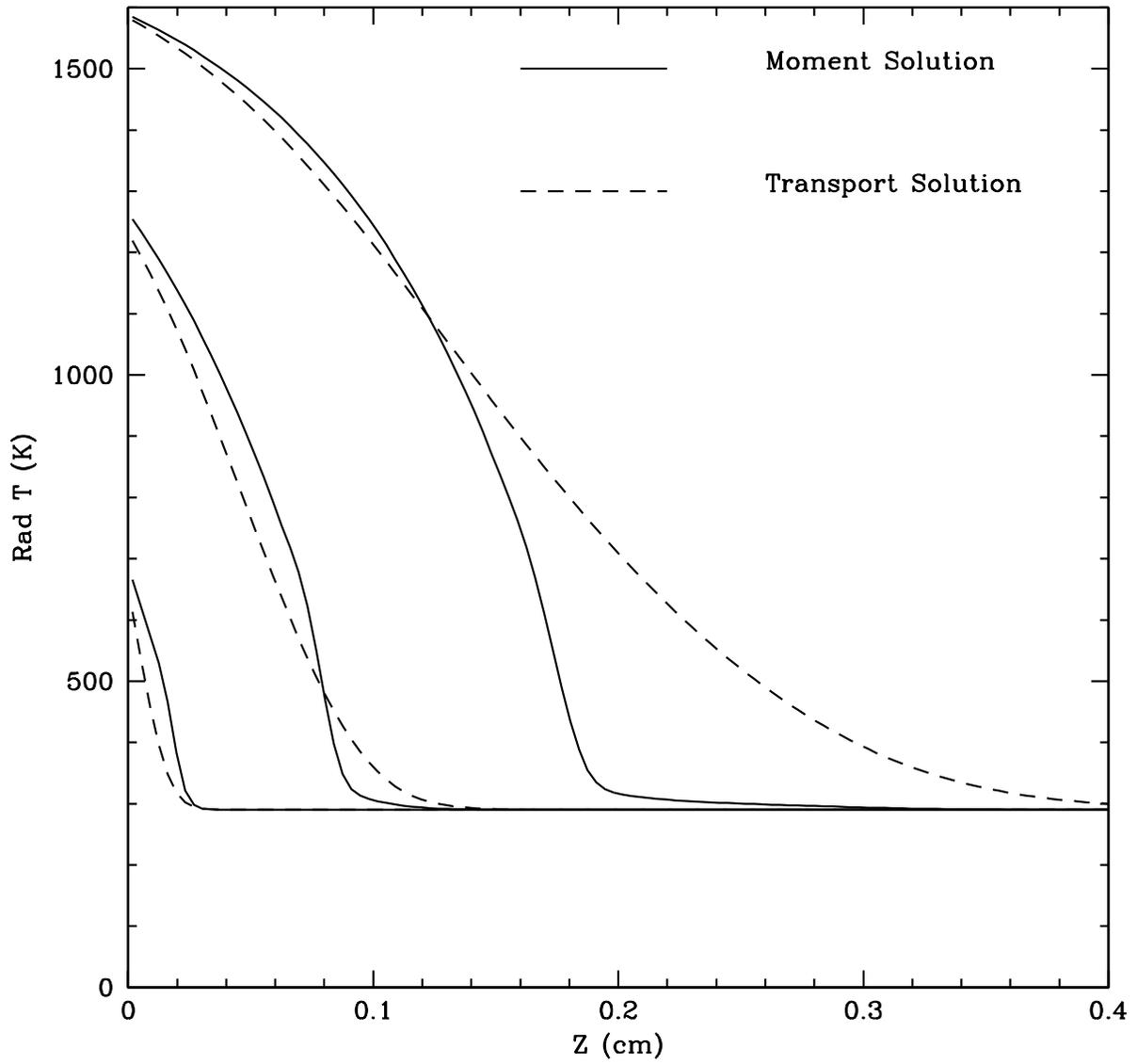}
\caption{As in figure~\ref{fig17}, but for a rapidly varying
radiation source.
\label{fig18}}
\end{figure}

We illustrate this behavior with comparing two streaming problems
computed with the full VTEF method.  The physical characteristics
of the two cases are identical and are taken from the cloud problem,
albeit with the dense cloud removed in favor of a uniform medium.
As with the cloud test, the illuminating source intensity increases
with a prescribed e-folding time.  In the case shown in figure
~\ref{fig17}, the e-folding time is equal to the light-crossing time
along the Z-axis; in figure~\ref{fig18} this value has been decreased
by an order of magnitude.  In both graphs we have plotted the radiation
temperature derived from the moment solution for the radiation energy
density, and compared it to that derived from an angular integration
of the specific intensity.  When the radiation source is illuminated
slowly, we see close agreement at all times between the two solutions.
When the radiation source is illuminated rapidly, there is a clear
disagreement between the two.  In particular, the profiles computed
from the transport solver begin to ``lead,'' to a significant degree,
those from the moment solution as the wavefront progresses.
The significance of this becomes apparent
when we realize that the Eddington tensor deviates from isotropy
everywhere that the specific intensity has risen from its ambient
value.  Once the Eddington tensor changes from its isotropic value,
the divergence terms in the flux equation become non-zero and generate
flux.  This means that fluxes from the moment solution at a given
point in space will become non-zero even before the radiation front
has overrun that point, clearly a non-physical result. 

As applied to the cloud problem, the consequences of these two
shortcomings are: (1) flux in the R-direction will leak radiation into
the shadowed region artificially, and (2) this behavior will occur
even before the radiation front has swept over the cloud, if the
radiation source is illuminated too rapidly.  As a practical matter,
this test problem needed to be designed with some care; attempting
to illuminate the source too quickly led to unrecoverable errors
in the radiation field evolution.  While there are phenomena
in astrophysics in which light-travel times may matter (e.g. the
emergence of radiation fronts from optically thick surfaces), problems abound
in which the radiation field may be treated as time independent, or
at worst quasistatic.  In this instance, our equations show good
internal consistency, and the dangers of pathological behavior are
greatly reduced.  Further, we note that considerable overall improvement
to the method will be obtained if our discrete solution is redesigned
to eliminate (or greatly reduce) the spurious generation of fluxes
at interface boundaries.  This problem is not unique to our method,
and alternative schemes for tensor diffusion problems have been
published in the radiation transport literature~\citep{mor98,mor01}.
With the support of the Department of Energy, we are now undertaking
to implement our VTEF equations with a higher-order spatial scheme
which satisfies the appropriate flux constraints.  Results of this
effort will be reported in future work.

The strengths of our method, in our view, are threefold.  With regard
to angular variations in the radiation intensity, the difference
between the VTEF and FLD solutions is fundamental and dramatic.  While
the VTEF solution shows room for quantitative improvement, the 
qualitative behavior is physically sensible, while that of the
FLD solution is not.  This result alone compels us to pursue the
VTEF approach further, with an eye toward relieving the diffusivity
issues identified above.  This algorithm is also particularly
suited to environments where matter and radiation are out
of equilibrium, a situation often encountered in strongly dynamic
environments.  The numerical scheme used to implement matter
radiation coupling has the further advantage of allowing an implicit
linear system solve for the radiation energy density alone, in
contrast to the two-variable coupled system documented in Paper III.
Finally, we note that our method is suited for large problems
distributed across parallel architectures.  This statement is even
truer for problems which require only a static transfer solution
for Eddington tensors (owing to a far lower memory requirement), or,
in the most favorable case, problems in which static Eddington tensors
can be computed analytically.  In this instance, the brunt of the
computational burden lies in the linear system solve for $E_{rad}$,
and is thus on the same order of the cost for an analogous solution
for FLD.  We therefore view this method as a topic worthy of continued
study, with regard to both algorithm development and astrophysical
applications.

\acknowledgments

This work owes an enormous debt to the previous effort of Prof.
James M. Stone, whose serial algorithm inspired the creation of the
parallel one detailed in this paper.  In addition, we are grateful to
Jim for two critical readings of the manuscript and for suggestions
that markedly improved its content.  Finally, we
gratefully acknowledge the continued support
and wise counsel from Dr. Frank Graziani at the Lawrence Livermore
National Laboratory.  This work was supported by DOE contract
W-7405-ENG-48.

\clearpage

\appendix

\section{The Radiation Moment Equation Linear System Matrix}
\label{matdoc}

Equation (\ref{rade4}), when written on a discrete mesh, leads
to a linear system for a solution vector consisting of the
discrete values of $E$ (the $E\subij$) everywhere on the moment
solution grid.  This matrix equation is represented schematically
by equation (\ref{radmat}), which shows that our system involves
a nine-band system.  The central $E$ value on the stencil ($E\subij$)
is coupled to $E\subimj$ and $E\subipj$ via the first subdiagonal
($DM1\subij$) and superdiagonal ($DP1\subij$), respectively.
Subdiagonals $DM4\subij$, $DM3\subij$, and $DM2\subij$ couple
solution elements $E\subimjm$, $E\subijm$, and $E\subipjm$,
respectively, and superdiagonals $DP2\subij$, $DP3\subij$, and
$DP4\subij$ couple elements $E\subimjp$, $E\subijp$, and $E\subipjp$.

Refering to equation (\ref{rade4}), we note that all three terms
on the LHS of (\ref{rade4}) contribute to $D00\subij$, but only
the second term of (\ref{rade4}) produces off-diagonal matrix
elements.  The third term of (\ref{rade4}) produces no off-diagonal
elements because it is viewed as product of the central $E\subij$
and the quantity $\nabla{\bf v} : \f$, where it is assumed that
$\nabla{\bf v}$ and $\f$ are already known.  Refering to the
definitions in \S~\ref{matradcoup}, we may therefore write for the main
diagonal:
\begin{equation}
D00\subij \ = \ \left(1 + {k_{2}(i,j) \over \Phi(i,j)}\right) \ + \
\Delta t (\nabla{\bf v} : \f)_{i,j} \ + \ T2_{00}(i,j),
\end{equation}
where $T2_{00}(i,j)$ are the contributions to $D00\subij$ from the
discretized second term in (\ref{rade4}).  The form of
$\nabla{\bf v} : \f$ has been documented in Paper III (albeit
with $\f$ replaced by $\p$), and will not be reproduced here.  The
remaining pieces of $D00\subij$ and all of the off-diagonal bands
are generated by the discrete form of $T2$, which we will now
document.

There are four steps to this documentation process.  The first step
is to properly express the required components of the tensor divergence
term, assuming 2-D cylindrical symmetry.  Considering our radiation
stress tensor, $\p$, the symmetry assumption implies that
$\p_{12}$, $\p_{21}$, $\p_{13}$, $\p_{31}$, $\p_{23}$, and $\p_{32}$
are all zero.  Regardless of symmetry, it is also true that the trace
of $\p$ is an invariant, and is identically equal to $E$.  This allows
us to write further that
\begin{equation}
\p_{33} \ = \ E - \p_{11} - \p_{22},
\end{equation}
or alternatively,
\begin{equation}
\f_{33} \ = \ 1 - \f_{11} - \f_{22},
\end{equation}
This relation was incorrectly expressed in the derivation of the
divergence terms documented in Paper III.

Making use of the general expressions in \citet{sto92a} for a tensor
divergence, and applying our conditions of symmetry and trace
invarience, we may write:
\begin{equation}
\left(\divp\right)^{(1)} \ = \ {1 \over g2g31}{\partial \over \partial
x_{1}}\left(g_{2}g_{31}\p_{11}\right) \ + \ {1 \over g_{2}}{\partial
\over \partial V_{2}}\left(g_{32}\p_{12}\right),
\end{equation}
and
\begin{equation}
\left(\divp\right)^{(2)} \ = \ g_{2}{\partial \over \partial V_{1}}
\left(g_{31}\p_{12}\right) \ + \ {1 \over g_{2}}{\partial \over
\partial V_{2}}\left(g_{32}\p_{22}\right) \ - \ {\left(E - \p_{11} -
\p_{22}\right) \over g_{2}g_{32}}{\partial g_{32} \over \partial x_{2}}.
\end{equation}

Step two of derivation is to employ the relation $\p = \f E$ and
discretize the tensor divergence terms, employing the same finite
difference conventions used in the ZEUS-2D code:
\begin{eqnarray}\label{divp1}
\left(\divp\right)^{(1)}\subij & \rightarrow &
{1 \over {\rm g2a\subi}~{\rm g31a\subi}}
\left[{{ {\rm g2b\subi g31b\subi}\f11\subij E\subij \ - \
{\rm g2b\subim g31b\subim}\f11\subimj E\subimj} \over {\rm dx1b\subi}}
\right] \nonumber \\
 & + & { {\rm g32a\subjp}\f12\subijp\eav\subijp \ - \ {\rm g32a\subj}
\f12\subij\eav\subij \over {\rm g2a\subi~dvl2a\subj}};
\end{eqnarray}
\begin{eqnarray}\label{divp2}
\left(\divp\right)^{(2)}\subij & \rightarrow &
\left[{ {\rm g32b\subj}\f22\subij E\subij \ - \ {\rm g32b\subjm}
\f22\subijm E\subijm } \over {\rm g2b\subi~dvl2b\subj}  \right] \nonumber \\
 & + & \left({\rm g2b\subi}\over{\rm dvl1a\subi}\right)
\left({\rm g31a\subip}\f12\subipj\eav\subipj \ - \ {\rm g31a\subi}
\f12\subij\eav\subij\right) \nonumber \\
 & - & \left(1 \over 2\right)\left(1 \over {\rm g2b\subi~g32a\subj}
\right)\left(\partial{\rm g32a}\subj\over\partial{\rm x2}\right)\times
\nonumber \\
&    & \left[E\subij  \left(1 - \f11\subij   - f22\subij  \right) +
      E\subijm1\left(1 - \f11\subijm1 - f22\subijm1\right) \right]
\end{eqnarray}
In (\ref{divp1}) and (\ref{divp2}), $\eav$ signifies an average
value of $E$ at the vertex of four neighboring zones; i.e.
\begin{equation}\label{ave}
\eav\subij \ = \ {1 \over 4}\left(E\subij \ + \ E\subimj \ + \
                                  E\subijm \ + \ E\subimjm \right)
\end{equation}

In step three, we write a discrete form of the ${\bf\nabla}\cdot
({\bf b}\divp)$, where {\bf b} is given by
\begin{equation}
{\bf b} \ = \ {c^{2}\Delta t \over \left(1 + c\Delta t\chif\right)}
\end{equation}
{\bf b} is written in boldface to indicate that it is dependent upon
the component of $\divp$ under consideration, owing to the variation
(in the general case) of $\chif$ with direction.
Employing our usual standard for writing vector divergences, we may
write
\begin{eqnarray}\label{divdiv}
{\bf\nabla}\cdot({\bf b}\divp) & \rightarrow &
{{\rm g2a\subip g31a\subip b^{(1)}\subipj}(\divp)\subipj ^{(1)} \ - \
 {\rm g2a\subi  g31a\subi  b^{(1)}\subij }(\divp)\subij  ^{(1)} \over
 {\rm dvl1a\subi} } \nonumber \\
 & + & {{\rm g32a\subjp b^{(2)}\subijp}(\divp)\subipj ^{(2)} \ - \
        {\rm g32a\subj  b^{(2)}\subij }(\divp)\subij  ^{(2)} \over
        {\rm g2b\subi~dvl2a\subj} }
\end{eqnarray}

The final step in the derivation of the matrix elements is the
substitution of expressions (\ref{divp1}) and (\ref{divp2}) into
(\ref{divdiv}).  All terms contributing to the main diagonal, for
example, are coefficients of the various terms containing $E\subij$.
Expressions for each of the sub- and superdiagonals are found by
collecting the coefficients of the appropriate discrete $E$ element
within the nine-point stencil.  We will not reproduce this algebra
here, but we document the final result.  Because the expressions
are complicated, we will subdivide result for each matrix band
in a manner reflective of the Fortran expressions used in the code.
We write:
\begin{mathletters}\label{bands}
\begin{eqnarray}
DM4\subij & = & C2\subij\cdot\bone\subi\cdot {\rm dp1}\subimjm \ + \
                C4\subij\cdot\btwo\subj\cdot {\rm dp2}\subimjm; \\
DM3\subij & = & C1\subij\cdot\bone\subip\cdot {\rm dp1p}\subijm \ + \
                C2\subij\cdot\bone\subi\cdot {\rm dp1}\subijm \nonumber \\
          & + & C4\subij\cdot\btwo\subj\cdot {\rm dp2}\subijm; \\
DM2\subij & = & C1\subij\cdot\bone\subip\cdot {\rm dp1p}\subipjm \ + \
                C4\subij\cdot\btwo\subj \cdot {\rm dp2} \subipjm; \\
DM1\subij & = & C2\subij\cdot\bone\subi \cdot {\rm dp1}\subimj \ + \
                C3\subij\cdot\btwo\subjp\cdot {\rm dp2p}\subimj \nonumber \\
          & + & C4\subij\cdot\btwo\subj \cdot {\rm dp2} \subimj; \\
D00\subij & = & C1\subij\cdot\bone\subip\cdot {\rm dp1p}\subij \ + \
                C2\subij\cdot\bone\subi \cdot {\rm dp1} \subij \nonumber \\
          & + & C3\subij\cdot\btwo\subjp\cdot {\rm dp2p}\subij \ + \
                C4\subij\cdot\btwo\subj \cdot {\rm dp2} \subij; \\
DP1\subij & = & C1\subij\cdot\bone\subip\cdot {\rm dp1p}\subipj \ + \
                C3\subij\cdot\btwo\subjp\cdot {\rm dp2p}\subipj \nonumber \\
          & + & C4\subij\cdot\btwo\subj \cdot {\rm dp2} \subipj; \\
DP2\subij & = & C2\subij\cdot\bone\subi \cdot {\rm dp1} \subimjp \ + \
                C3\subij\cdot\btwo\subjp\cdot {\rm dp2p}\subimjp; \\
DP3\subij & = & C1\subij\cdot\bone\subip\cdot {\rm dp1p}\subijp \ + \
                C2\subij\cdot\bone\subi \cdot {\rm dp1} \subijp \nonumber \\
          & + & C3\subij\cdot\btwo\subjp\cdot {\rm dp2p}\subijp; \\
DP4\subij & = & C1\subij\cdot\bone\subip\cdot {\rm dp1p}\subipjp \ + \
                C3\subij\cdot\btwo\subjp\cdot {\rm dp2p}\subipjp;
\end{eqnarray}
\end{mathletters}
where
\begin{eqnarray}\label{coeffs}
C1\subij & = & -\Delta t {{\rm g2a\subip}~{\rm g31a\subip} \over
                       {\rm dvl1a\subi}}; \\
C2\subij & = & \ \ \Delta t {{\rm g2a\subi }~{\rm g31a\subi } \over
                  {\rm dvl1a\subi}}; \\
C3\subij & = & -\Delta t {{\rm g32a\subjp} \over {\rm g2b\subi}~
                  {\rm dvl2a\subj}}; \\
C4\subij & = &  \ \ \Delta t {{\rm g32a\subj } \over {\rm g2b\subi}~
                  {\rm dvl2a\subj}}.
\end{eqnarray}

In (\ref{bands}), ``dp1'' is shorthand for the coefficients of
a particular element of $E$ in the expression for $(\divp)\subij^{(1)}$.
The element of $E$ in question is indicated by the subscripts on
dp1.  There are analogous relationships between ``dp1p'' and
$(\divp)\subipj^{(1)}$, ``dp2'' and $(\divp)\subij ^{(2)}$, and
``dp2p'' and $(\divp)\subijp^{(2)}$.  Save for modifications due to
applied boundary conditions, we complete our documentation of the
matrix elements by listing the values for the four
``dp$\{n\}$'' quantities at the required $E$ index values.  For
$(\divp)^{(1)}$ at $(i,j)$, we have:
\begin{eqnarray}
{\rm dp1}\subimjm & = & -\left(1 \over 4\right)
                      { {\rm g32a}\subj\f12\subij \over
                      {\rm g2a}\subi~{\rm dvl2a}\subj }; \\
{\rm dp1}\subijm  & = & -\left(1 \over 4\right)
                      { {\rm g32a}\subj\f12\subij \over
                      {\rm g2a}\subi~{\rm dvl2a}\subj };
\end{eqnarray}
\begin{eqnarray}
{\rm dp1}\subimj  & = &
                    -{ {\rm g2b}\subim{\rm g31b}\subim\f11\subimj \over
          {\rm g2a}\subi~{\rm g31a}\subi~{\rm dx1b}\subi } \nonumber \\
 & + & \left(1 \over 4\right){ {{\rm g32a}\subjp\f12\subijp \ - \
                               {\rm g32a}\subj \f12\subij}
                         \over {\rm g2a}\subi~{\rm dvl2a}\subj };
\end{eqnarray}
\begin{eqnarray}
{\rm dp1}\subij   & = & { {\rm g2b}\subi{\rm g31b}\subi\f11\subij \over
                    {\rm g2a}\subi~{\rm g31a}\subi~{\rm dx1b}\subi }
 \ + \ \left(1 \over 4\right){ {{\rm g32a}\subjp\f12\subijp \ - \
                               {\rm g32a}\subj \f12\subij}
                         \over {\rm g2a}\subi~{\rm dvl2a}\subj }; \\
{\rm dp1}\subimjp & = & \left(1 \over 4\right)
                      { {\rm g32a}\subjp\f12\subijp \over
                      {\rm g2a}\subi~{\rm dvl2a}\subj }; \\
{\rm dp1}\subijp  & = & \left(1 \over 4\right)
                      { {\rm g32a}\subjp\f12\subijp \over
                      {\rm g2a}\subi~{\rm dvl2a}\subj }.
\end{eqnarray}
For $(\divp)^{(1)}$ at $(i+1,j)$, we have:
\begin{eqnarray}
{\rm dp1p}\subijm & = & -\left(1 \over 4\right)
                      { {\rm g32a}\subj\f12\subipj \over
                      {\rm g2a}\subip~{\rm dvl2a}\subj }; \\
{\rm dp1p}\subipjm  & = & -\left(1 \over 4\right)
                      { {\rm g32a}\subj\f12\subipj \over
                      {\rm g2a}\subip~{\rm dvl2a}\subj };
\end{eqnarray}
\begin{eqnarray}
{\rm dp1p}\subij  & = & -{{\rm g2b}\subi{\rm g31b}\subi\f11\subij\over
       {\rm g2a}\subip~{\rm g31a}\subip~{\rm dx1b}\subip} \nonumber \\
  & + & \left(1 \over 4\right){ {{\rm g32a}\subjp\f12\subipjp \ - \
                          {\rm g32a}\subj \f12\subipj}
                        \over {\rm g2a}\subip~{\rm dvl2a}\subj };
\end{eqnarray}
\begin{eqnarray}
{\rm dp1p}\subipj   & = & {{\rm g2b}\subip{\rm g31b}\subi\f11\subipj
   \over {\rm g2a}\subip~{\rm g31a}\subip~{\rm dx1b}\subip} \nonumber \\
  & + & \left(1 \over 4\right){ {{\rm g32a}\subjp\f12\subipjp \ - \
                          {\rm g32a}\subj \f12\subipj}
                        \over {\rm g2a}\subip~{\rm dvl2a}\subj };
\end{eqnarray}
\begin{eqnarray}
{\rm dp1p}\subijp & = & \left(1 \over 4\right)
                      { {\rm g32a}\subjp\f12\subipjp \over
                      {\rm g2a}\subip~{\rm dvl2a}\subj }; \\
{\rm dp1p}\subipjp  & = & \left(1 \over 4\right)
                      {{\rm g32a}\subjp\f12\subipjp\over
                      {\rm g2a}\subip~{\rm dvl2a}\subj }.
\end{eqnarray}
For $(\divp)^{(2)}$ at $(i,j)$, we have:
\begin{equation}
{\rm dp2}\subimjm \ = \ - \left(1 \over 4\right)
                      { {\rm g2b}\subi{\rm g31a}\subi
                      \f12\subij \over {\rm dvl1a}\subi };
\end{equation}
\begin{eqnarray}
{\rm dp2}\subijm  & = & - { {\rm g32b}\subjm\f22\subijm \over
                        {\rm g2b}\subi~{\rm dvl2b}\subj }
            \ - \ \left(1 \over 2\right)
                   {1 \over {\rm g2b}\subi~{\rm g32a}\subj}
                   {\partial {\rm g32a}\subj \over \partial {\rm x2}}
       \left(1 - \f11\subijm - \f22\subijm\right) \nonumber \\
          & + & \left(1 \over 4\right){ {\rm g2b}\subi
                \over {\rm dvl1a}\subi }
              \left({\rm g31a}\subip\f12\subipj \ - \
                    {\rm g31a}\subi \f12\subij\right);
\end{eqnarray}
\begin{eqnarray}
{\rm dp2}\subipjm & = & \left(1 \over 4\right)
                      { {\rm g2b}\subi{\rm g31a}\subip
                      \f12\subipj \over {\rm dvl1a}\subi }; \\
{\rm dp2}\subimj  & = & -\left(1 \over 4\right)
                      { {\rm g2b}\subi{\rm g31a}\subi
                      \f12\subij \over {\rm dvl1a}\subi };
\end{eqnarray}
\begin{eqnarray}
{\rm dp2}\subij   & = & { {\rm g32b}\subj\f22\subij \over
                        {\rm g2b}\subi~{\rm dvl2b}\subj }
 \ - \ \left(1 \over 2\right){1 \over {\rm g2b}\subi~{\rm g32a}\subj}
                   {\partial {\rm g32a}\subj \over \partial {\rm x2}}
          \left(1 - \f11\subij - \f22\subij\right) \nonumber \\
                  & + & \left(1 \over 4\right){ {\rm g2b}\subi
                      \over {\rm dvl1a}\subi }
                  \left({\rm g31a}\subip\f12\subipj \ - \
                        {\rm g31a}\subi\f12\subij\right);
\end{eqnarray}
\begin{equation}
{\rm dp2}\subipj  \ = \ \left(1 \over 4\right)
                      { {\rm g2b}\subi{\rm g31a}\subip
                      \f12\subipj \over {\rm dvl1a}\subi }.
\end{equation}
For $(\divp)^{(2)}$ at $(i,j+1)$, we have:
\begin{equation}
{\rm dp2p}\subimj \ = \ - \left(1 \over 4\right)
                      { {\rm g2b}\subi{\rm g31a}\subi
                      \f12\subij\over {\rm dvl1a}\subi };
\end{equation}
\begin{eqnarray}
{\rm dp2p}\subij  & = & - { {\rm g32b}\subj \f22\subij  \over
                        {\rm g2b}\subi~{\rm dvl2b}\subjp}
 \ - \ \left(1 \over 2\right){1 \over {\rm g2b}\subi~{\rm g32a}\subjp}
                   {\partial {\rm g32a}\subjp\over \partial {\rm x2}}
           \left(1 - \f11\subij - \f22\subij\right) \nonumber \\
         & + & \left(1 \over 4\right){ {\rm g2b}\subi
                       \over {\rm dvl1a}\subi }
                       \left({\rm g31a}\subip\f12\subipjp \ - \
                             {\rm g31a}\subi\f12\subijp\right);
\end{eqnarray}
\begin{eqnarray}
{\rm dp2p}\subipj & = & \left(1 \over 4\right)
                       { {\rm g2b}\subi{\rm g31a}\subip
                      \f12\subipjp \over {\rm dvl1a}\subi }; \\
{\rm dp2p}\subimjp & = & - \left(1 \over 4\right)
                      { {\rm g2b}\subi{\rm g31a}\subi
                      \f12\subijp \over {\rm dvl1a}\subi };
\end{eqnarray}
\begin{eqnarray}
{\rm dp2p}\subijp  & = & { {\rm g32b}\subjp\f22\subijp \over
                        {\rm g2b}\subi~{\rm dvl2b}\subjp}
 \ - \ \left(1 \over 2\right){1 \over {\rm g2b}\subi~{\rm g32a}\subjp}
                   {\partial {\rm g32a}\subjp \over \partial {\rm x2}}
       \left(1 - \f11\subijp - \f22\subijp\right) \nonumber \\
     & + & \left(1 \over 4\right){ {\rm g2b}\subi
                       \over {\rm dvl1a}\subi }
                  \left({\rm g31a}\subip\f12\subipjp \ - \
                        {\rm g31a}\subi\f12\subijp\right);
\end{eqnarray}
\begin{eqnarray}
{\rm dp2p}\subipjp & = & \left(1 \over 4\right)
                      { {\rm g2b}\subi{\rm g31a}\subip
                      \f12\subipjp \over {\rm dvl1a}\subi }.
\end{eqnarray}

\clearpage

\end{document}